# Centre-of-mass separation in quantum mechanics: Implications for the many-body treatment in quantum chemistry and solid state physics

## Michal Svrček

---

**Abstract**

We address the question to what extent the centre-of-mass (COM) separation can change our view of the many-body problem in quantum chemistry and solid state physics. It was shown that the many-body treatment based on the electron-vibrational Hamiltonian is fundamentally inconsistent with the Born-Handy ansatz so that such a treatment can never respect the COM problem. Born-Oppenheimer (B-O) approximation reveals some secret: it is a limit case where the degrees of freedom can be treated in a classical way. Beyond the B-O approximation they are inseparable in principle. The unique covariant description of all equations with respect to individual degrees of freedom leads to new types of interaction: besides the known vibronic (electron-phonon) one the rotonic (electron-roton) and translonic (electron-translon) interactions arise. We have proved that due to the COM problem only the hypervibrations (hyperphonons, i.e. phonons + rotons + translons) have true physical meaning in molecules and crystals; nevertheless, the use of pure vibrations (phonons) is justified only in the adiabatic systems. This fact calls for the total revision of our contemporary knowledge of all non-adiabatic effects, especially the Jahn-Teller effect and superconductivity. The vibronic coupling is responsible only for removing of electron (quasi)degeneracies but for the explanation of symmetry breaking and forming of structure the rotonic and translonic coupling is necessary.

## 1. Introduction

As it is well known, the true many body methods including the Feynman diagrammatic technique, developed in the elementary particle physics, were transferred in the solid state physics many years ago. Their introduction in quantum chemistry followed later, but only on the electronic level. So the question has arisen: Is it possible to construct the full quantum chemical electron-vibrational Hamiltonian in the second quantization formalism?

The answer is no. After the many years spent on the attempt to construct the ideal representation by means of quasiparticle transformations (the equivalent of Fröhlich type unitary transformations), all the variants, either the adiabatic representation or the nonadiabatic one, have failed. The reason lies deeper than one would initially imagine.

Two scientific disciplines, quantum chemistry and solid state physics, were build on the mathematical simplification of Schrödinger equation, known as the Born-Oppenheimer (B-O) approximation [1]. It is not only the basis of almost all quantum chemical calculations, but also of the very concept of molecular structure [2].

There are two main contemporary trends in quantum chemistry that put a question mark over B-O approximation and its role in the definition of structure: the theories based on the inclusion of the centre-of-mass (COM) problem and the applications of the Jahn-Teller (J-T) effect.

There is no problem to include the COM problem in the calculations of atoms but its implementation in molecules is very complicated. Monkhorst [3] proposed a model of molecular atoms for this purpose. The practical profit of this approach is limited only to the smallest molecules as we can see in later works of Cafiero and Adamowitz [4], based on Monkhorst's ideas: "We have the analogue of the nucleus with the heavy particle at the center of the internal coordinate system, and we have the analogues of electrons in the internal particles. The main difference between this model and an atom is that the internal particles in an atom are all electrons and in the "molecular atom" or "atomic molecule" the internal particles may be both electrons and nuclei (or, as we should more correctly say, pseudoparticles resembling the electron and the nuclei). Formally this difference manifests itself in the effective masses of the pseudoparticles and in the way the permutational symmetry is implemented in the wave function" [4].

This article contains an interesting note about the structure: "While molecular structure is a central concept in chemistry and physics, it should be remembered that for an isolated gas phase molecule in field-free space the most information that can be acquired is the average values of structural parameters (i.e., bond distances and angles). This point becomes apparent when molecular calculations are done without the B-O approximation - an almost universal approximation in quantum chemistry. While this approximation is extremely useful and has largely defined the terminology of modern spectroscopy, it also hides some simple quantum mechanical truths about the systems we study."

Moreover, the very limited applicability of the COM separation is not the only one problem; another problem arises from the introduction of degrees of freedom for molecules, which are absent in atoms. Therefore Kutzelnigg [5] writes: "The adiabatic approximation with COM separation is good for atoms, because it takes care that the electrons participate in the COM motion. However, it is unbalanced for molecules, because it favours the COM motion with respect to other motions dominated by the nuclei, such as rotation and vibration, where the participation of the electrons is less trivial anyway than in the COM motion. The partial participation of the electrons in these motions is ignored in the adiabatic approximation both with and without COM separation."

That is the reason why Kutzelnigg finally prefers the pragmatic Born-Handy ansatz [6] when he has proved its full equivalency to the COM separation: "Handy and co-workers have never claimed to have invented the ansatz referred to here as the "Born-Handy ansatz", but they certainly convinced a large audience that this ansatz is of enormous practical value, even if it has not been completely obvious why it leads to correct results. Handy and co-workers realized that the difficulties with the traditional approach come from the separation of the COM motion (and the need to define internal coordinates after this separation has been made). They therefore decided to renounce the separation."

Surely, this is a significant progress since the applicability of the Born-Handy ansatz as a full replacement of the COM separation is not restricted by the size of the investigated system. But unfortunately, its applicability is limited only to adiabatic systems. Born-Handy ansatz is able to yield only the adiabatic corrections to the B-O results, and beyond this ansatz spreads out the enigmatic region denoted as the break-down of B-O approximation. Therefore one of the main goal of this work is the finding of the "extended Born-Handy formula", valid in the adiabatic limit as well as beyond it.

The most important consequence of the breakdown of B-O approximation is indisputably the Jahn-Teller (J-T) effect [7] where the structure defined on the basis of this approximation does not hold any more. The important role of the J-T effect is emphasized in the Bersuker's book [8]: "Moreover, since the J-T effect has been shown to be the only source of spontaneous distortion of high-symmetry configurations, we come to the conclusion that the J-T effect is a unique mechanism of all the symmetry breakings in condensed matter."

The fact, that in solid state physics the problem of the definition of crystallic structure is mostly ignored and only the B-O structure is supposed, is often criticised by scientists devoting to the J-T effect. Especially the superconductors are evidently the non-adiabatic problem but the impact on the crystallic structure is reflected neither in the Fröhlich Hamiltonian [9-10] nor in the BCS theory [11]. Bersuker writes about the implications of the J-T effect in the superconductivity problem:

"An illustration of the JT approach to electron–phonon coupling in solids may be found in the modern attempts to explain the origin of high-temperature superconductivity (HTSC). Experimental data show that the electron–phonon interaction is essential in this phenomenon… The existing BCS theory of superconductivity takes into account the electron–phonon interaction "in general" as an interaction of the electrons with the "bath of phonons" without detailed analysis of the local aspects of this interaction leading to the J-T effect. For broad-band metals with widely delocalized electrons the J-T electron–phonon coupling is weak and the J-T effect may be ignored. This is why the BCS theory explains the origin of superconductivity at low temperatures without taking into account the J-T effect. For narrower bands (which are characteristic of systems with HTSC) the J-T effect becomes significant, and the application of the achievements of the J-T effect theory to the HTSC problem seems to be most appropriate. This is indeed the subject of most current attempts to treat the HTSC yielding reasonable (reassuring) results."

At this stage the J-T approach in superconductivity problem is still questionable from two reasons: first, the J-T approach is not formulated in the many-body form as the Fröhlich Hamiltonian is, and second, this approach ignores the COM problem, but Fröhlich Hamiltonian and all theories based on it, including the BCS one, suffer from the same ignoring, too.

What is extremely desirable now: There has to exist some unifying way so that both trends – Born-Handy alternative of the COM separation and the J-T approach could shake hands. Maybe it seems on first sight as a utopian requirement, because the Born-Handy formula respects the COM problem but is only valid within the adiabatic limit, and on the other hand, the J-T approach solves the non-adiabatic problems but does not respect the COM problem.

As we demonstrate in this paper, this goal is achievable through the revision of our knowledge of the true many-body second quantization formalism and the reopening of the old question, what forms the molecular and crystallic structure when the B-O approximation breaks down. Moreover, this revision fulfils till now unsatisfied personal views of many prominent scientists: It will be confirmed the Löwdin's idea, that the electron quasidegeneracies are unnatural, and the Fröhlich's idea, that the superconductivity has to be of one-particle origin. Although the Fröhlich Hamiltonian was incorporated in the BCS theory, Fröhlich never accepted this two-particle theory from the above mentioned reason.

In the first phase of the development of quantum mechanics it was believed, when Schrödinger equation was formulated for any number of nuclei and electrons, so that the calculation of molecules and crystals is a pure application in a similar way as the calculation of the motion of complex objects by means of Newton mechanics. This fact of "boring automatic application of Schrödinger equation" caused the outflow of many prominent scientists into another branches - nuclear physics, elementary particle physics, and later solid state physics. Quantum chemistry was underestimated for many years, although it should have been the cradle of the paradigmatic background of the true-many body treatment, but unfortunately never was. The region slightly beyond the hydrogen atom, the simplest molecules, especially the hydrogen molecule plays the decisive role in this aspect.

The solid state physicists were inspired with the quantum field electron-photon interaction, and jumping over nuclei, over atoms, over molecules, transferred these ideas into solid state physics in the form of many-body treatment based on the electron-phonon interaction, in spite of the fact that the limits of its validity were never investigated, especially with the relationship to the COM problem.

The principal fundaments of the approach I am going to present were laid already in my PhD thesis in 1986 [12] and in an unpublished work from 1988 [13]. In a full analogy with the solid state electron-phonon interaction treatment, I started to build the similar apparatus for quantum chemistry, the second quantization quasiparticle concept of electron-vibrational

Hamiltonian. It was not an easy task at all as I expected on the beginning. First the crude representation was formulated, then the adiabatic, and finally the nonadiabatic one. As I have later recognized, the quasiparticle transformations leading to individual representations were the full quantum chemistry equivalent of Fröhlich transformations used in solid state physics.

Most of the equations and concepts were subsequently published in a series of papers [14-21]. However, some essential ingredients were missing in my original formulation, namely the inclusion of the COM problem. In particular, in the works [14-17] the molecular adiabatic corrections were wrong derived without having regard for Born-Handy ansatz. Similarly, the concept of superconductivity presented in [18-21] was restricted to another solution of Fröhlich equation for the ground state, where the structural instability was supposed. But the most important term - the trigger responsible for this structural instability - was missing there. Only if these deficiencies are remedied, a full picture of a truly COM compatible many-body treatment of nuclei and electrons on the same footing emerges.

I was a little bit confused in 1998 when Biskupič during the numerical tests performing on H2, HD and D2 molecules gained from the equations of my original formulation, where the COM problem was not considered, only 20 % contribution of the total groundstate adiabatic correction in comparison with the standard methods based on the Born-Handy formula [22]. I blamed him first for some bug in programming but soon it was evident that this was my fault. I tried to convert Born-Handy formula in the form compatible with the Coupled Perturbed Hartree-Fock (CPHF) [23,24] notation just as my theory was expressed in the CPHF form, so that both - Born-Handy ansatz and my theory could be directly compared.

At the first moment it looked like some curiosity: hydrogen molecule does not move, does not rotate, and nevertheless the final Born-Handy formula contains contributions from vibrational as well as from translational and rotational degrees of freedom, whereas my theory based on the quantization of electron-vibrational Hamiltonian contains solely the contributions from the vibrational degrees. As it will be shown below, this has a profound significance for all systems and phenomena beyond the Born-Oppenheimer approximation. Moreover, the interpretation of degrees of freedom in quantum mechanics will be discussed, since the presented theory gives them quite different meaning than in classical mechanics.

In this work an emphasis will be laid on those steps, which deviate from an approach where the COM problem is ignored. I start with the summarizing of the conversion of Born-Handy ansatz into the CPHF compatible form [22], and will continue with the consequences of this important result for the correct quantization of the total Hamiltonian. Then the downward rederivation of Born-Handy formula from the new theory will follow, and finally the consequences of the new theory for the Jahn-Teller effect, conductors and superconductors will be presented.

## 2. Conversion of the Born-Handy formula in the CPHF compatible form

Let us start with the Born-Handy ansatz [6] for the groundstate electron wavefunction $\psi_0(\vec{R})$ where $\vec{R}$ represents nuclear coordinates. The adiabatic correction $\Delta E_0$ to the groundstate is expressed as a mean value of the nuclear kinetic operator $T_N$ [22],

$$\Delta E_0 = \left\langle \psi_0(\vec{R}) \middle| T_N \middle| \psi_0(\vec{R}) \right\rangle_{R_0}$$

$$= \sum_{i\alpha} \frac{\hbar^2}{2M_i} \left\langle \frac{\partial \psi_0(\vec{R})}{\partial R_{i\alpha}} \middle| \frac{\partial \psi_0(\vec{R})}{\partial R_{i\alpha}} \right\rangle_{R_0} \tag{2.1}$$

where after the integration per partes indexes i denote nuclei, $\alpha$ Cartesian coordinates and $M_i$ nuclear mass. In the adiabatic case the N-electron function $\psi_0(\vec{R})$ can be expanded as a single Slater determinant though the one-electron functions $\varphi_I(\vec{R})$:

$$\psi_0(\vec{R}) = \frac{1}{\sqrt{N!}} \left\| \prod_I^N \varphi_I(\vec{R}) \right\| \tag{2.2}$$

In this whole work we will use the following notation for spinorbitals: I, J, K, L – occupied; A, B, C, D – virtual (unoccupied); P, Q, R, S – arbitrary ones. Substituting (2.2) for $\psi_0(\vec{R})$ in (2.1) we get:

$$\Delta E_0 = \sum_{i\alpha} \frac{\hbar^2}{2M_i} \left( \sum_I \left\langle \frac{\partial \varphi_I}{\partial R_{i\alpha}} \middle| \frac{\partial \varphi_I}{\partial R_{i\alpha}} \right\rangle_{R_0} \right.$$

$$\left. - \sum_{I \neq J} \left\langle \varphi_J \middle| \frac{\partial \varphi_J}{\partial R_{i\alpha}} \right\rangle \left\langle \frac{\partial \varphi_I}{\partial R_{i\alpha}} \middle| \varphi_I \right\rangle_{R_0} - \sum_{I \neq J} \left\langle \varphi_J \middle| \frac{\partial \varphi_I}{\partial R_{i\alpha}} \right\rangle \left\langle \frac{\partial \varphi_I}{\partial R_{i\alpha}} \middle| \varphi_J \right\rangle_{R_0} \right) \tag{2.3}$$

The function $\varphi_P(\vec{R})$ can be expanded through the nuclear coordinates dependent coefficients $c_{PQ}(\vec{R})$ and the orthonormal set of one-electron wavefunctions defined in the equilibrium position $R_0$.

$$\varphi_P(\vec{R}) = \sum_Q c_{PQ}(\vec{R}) \varphi_Q(R_0) \tag{2.4}$$

The equation (2.3) now reads:

$$\Delta E_0 = \sum_{i\alpha} \frac{\hbar^2}{2M_i} \left( \sum_{QI} \left| c_{QI}^{i\alpha} \right|^2 - \sum_{I \neq J} c_{IJ}^{i\alpha} c_{II}^{i\alpha *} - \sum_{I \neq J} \left| c_{IJ}^{i\alpha} \right|^2 \right) \tag{2.5}$$

where indices $i\alpha$ are related to first derivatives. Using the calibration for the diagonal coefficients in accordance with the CHPF formulation [22-24]

$$c_{II}^{i\alpha} = 0 \tag{2.6}$$

we arrive the very simple CHPF form of Born-Handy formula for the groundstate:

$$\Delta E_0 = \sum_{AIi\alpha} \frac{\hbar^2}{2M_i} \left| c_{AI}^{i\alpha} \right|^2 \tag{2.7}$$

This is the expression based on real coordinates $R_{i\alpha}$. But the analogical expression based on normal ones $Q_r$ will be much more interesting. Therefore we introduce, again in accordance with the CHPF formulation, the expansion coefficients $c^r_{PQ}$.

$$c^r_{PQ} = \sum_{i\alpha} c^{i\alpha}_{PQ} \frac{\partial R_{i\alpha}}{\partial Q_r} = \sum_{i\alpha} c^{i\alpha}_{PQ} \alpha^r_{i\alpha} \qquad (2.8)$$

In order to substitute for $c^{i\alpha}_{PQ}$ in (2.7), we need to know the inverse matrix $\beta^r_{i\alpha}$. Since it holds

$$\alpha^+ \beta = I \qquad (2.9)$$

(2.7) can be rewritten as

$$\Delta E_0 = \sum_{Alrsi\alpha} \frac{\hbar^2}{2M_i} c^r_{Al} c^{s\,*}_{Al} \beta^{r\,*}_{i\alpha} \beta^s_{i\alpha} \qquad (2.10)$$

It is very important to say now, that the summation in (2.10) must be performed over all degrees of freedom, i.e. 3N, including 2 or 3 rotational and 3 translational ones! From the equiparticle theorem we know that the potential as well as the kinetic energy, standardly defined in accordance with the CPHF theory, contribute the same values – the halves of vibrational energy. But whereas diagonalizing the potential energy we get 3N - 5 or 3N - 6 nonzero values for vibrational modes and 5 or 6 zero values, the same procedure applied on the kinetic energy gives us 3N nonzero values, aside the vibrational modes we have 2 or 3 nonzero values for rotational modes and 3 nonzero values for translational ones.

$$\alpha^+ E_{pot} \alpha = \left\{ \frac{1}{2}\hbar\omega, 0, 0 \right\} \qquad (2.11)$$

$$\beta^+ E_{kin} \beta = \left\{ \frac{1}{2}\hbar\omega, \rho, \tau \right\} = \sum_{i\alpha} \frac{\hbar^2}{4M_i} \beta^{\,*}_{i\alpha} \beta_{i\alpha} \qquad (2.12)$$

With the help of (2.12) we can simplify the term inside of the equation (2.10),

$$\sum_{i\alpha} \frac{\hbar^2}{2M_i} \beta^{r\,*}_{i\alpha} \beta^s_{i\alpha} = \left\{ \hbar\omega_r \big|_{r\in V}, 2\rho_r \big|_{r\in R}, 2\tau_r \big|_{r\in T} \right\} \delta_{rs} \qquad (2.13)$$

where $\omega_r$ represent the vibrational frequencies, whereas the $\rho_r$ and $\tau_r$ represent the quanta of energy of the rotational and vibrational modes, extracted from the same secular equation for the kinetic energy (2.12) as the vibrational frequencies.

$$\Delta E_0 = 2\sum_{Al} \left( \sum_{r\in V} \frac{1}{2}\hbar\omega_r + \sum_{r\in R} \rho_r + \sum_{r\in T} \tau_r \right) \left| c^r_{Al} \right|^2 \qquad (2.14)$$

The final form of the Born-Handy formula consists of three terms: The first one represents the electron-vibrational interaction. I will not present the numerical details for $H_2$, HD and $D_2$ molecules here, it can be found in our previous work. The most important resulting fact is that the electron-vibrational Hamiltonian is totally insufficient for the description of the adiabatic correction to the molecular groundstates; its contribution differs almost in one decimal place from the real values acquired from the Born-Handy formula. In the case of concrete examples - $H_2$, HD and $D_2$ molecules - the first term contributes only with cca 20 % of the total value. The dominant rest - 80 % of the total contribution - descends of the electron-translational and electron-rotational interaction [22]. This interesting effect occurs on one-particle level, and it justifies the use of one-determinant expansion of the wave function (2.2). Of course, we can calculate the corrections beyond the Hartree-Fock approximation by means of many-body perturbation theory, as it was done in our work [22], but at this moment it is irrelevant to further considerations.

The Born-Handy ansatz [6] was verified on simple molecular systems many times in the last years [25-27], and especially interesting is the comparison of this simple pragmatic ansatz with the rigorous methods based on the separation of the centre-of-mass motion where one gets rather complicated expressions in terms of relative coordinates in a molecule-fixed frame. Kutzelnigg proved the validity of the Born-Handy ansatz by means of centre-of-mass analysis [5]. Now we can ask: what about the case of the break-down of adiabatic approximation? If the adiabatic case beyond the B-O approximation is a centre-of-mass problem in general, so much the more the break-down case is the same problem. But where is any reference to centre-of-mass problem in papers and books about e.g. the Jahn-Teller effect or superconductivity?

The centre-of-mass methods are based on the introduction of relative coordinates and relative effective masses and are complicated. Their applicability in the case of the break-down of adiabatic approximation is therefore very questionable. Looking at the CPHF reformulation (2.14) we must agree that this equation is very simple. Equation (2.14) bounds together the $3N - 6(5)$ vibration modes with 3(2) rotation and 3 translation ones. Therefore the new theory must be strictly built on the covariant notation regarding the individual degrees of freedom. We will call them $3N$-dimensional hypervibrations in quantum chemistry and hyperphonons in solid state physics.

## 3. Reconstruction of the total Hamiltonian in the second quantization formalism

At the opening of this topic we point out some important remarks about the cross-platform notation used in this paper. As one can see in the previous chapter we have used expansion coefficients $c_{PQ}$ of the one-electron wavefunctions and eigenvectors $\alpha_{i\alpha}^r$ / $\beta_{i\alpha}^r$ of the secular equations for the potential / kinetic oscillator energy defined on the set of complex numbers. In quantum chemistry this is irrelevant, all the mentioned coefficients can be real, but in solid state physics the complex number notation is necessary. In a similar way we will use the cross-platform notation for coordinate and momentum oscillator operators, namely $B_r = b_r + b_r^+$ and $\tilde{B}_r = b_r - b_r^+$. In quantum chemistry simply holds $r = \check{r}$. In solid state physics we assume that for any vibrational mode $r$ there exists corresponding mode $\check{r}$ fulfilling the identity $\omega_r = \omega_{\bar{r}}$. Comparing this notation with the usual solid state notation the simple transition $P \to \mathbf{k}, \sigma, r \to \mathbf{q}, \bar{r} \to \mathbf{-q}$ is supposed.

The Born-Oppenheimer approximation leads in the final stage to the system of independent harmonic oscillators. The B-O vibrational Hamiltonian $H_{BO}$ reads

$$H_{BO} = \sum_{r \in V} \hbar \omega_r \left( b_r^+ b_r + \tfrac{1}{2} \right) = \frac{1}{4} \sum_{r \in V} \hbar \omega_r \left( B_r^+ B_r + \tilde{B}_r^+ \tilde{B}_r \right) \tag{3.1}$$

where $r$ represents only the vibrational modes.

First we proceed to the introduction of hypervibrational modes. In accordance with secular equations (2.11) and (2.12) we get the following second quantization form for the potential and kinetic energies:

$$E_{pot} = \frac{1}{4} \sum_{r \in V} \hbar \omega_r B_r^+ B_r \tag{3.2}$$

$$E_{kin} = \frac{1}{4} \left( \sum_{r \in V} \hbar \omega_r + 2 \sum_{r \in R} \rho_r + 2 \sum_{r \in T} \tau_r \right) \tilde{B}_r^+ \tilde{B}_r \tag{3.3}$$

Let us denote the hypervibrational Hamiltonian as $H_B$.

$$H_B = E_{kin}(\tilde{B}) + E_{pot}(B) \tag{3.4}$$

In order to get the covariant form of the Hamiltonian $H_B$, we will define the hypervibrational double-vector:

$$\boldsymbol{\omega} = \begin{pmatrix} \omega_r \\ \tilde{\omega}_r \end{pmatrix} = \begin{pmatrix} \omega_r & 0 & 0 \\ \omega_r & \dfrac{2}{\hbar} \rho_r & \dfrac{2}{\hbar} \tau_r \end{pmatrix} \tag{3.5}$$

This notation leads to the fully covariant expression for the Hamiltonian $H_B$ with respect of all 3N hypervibrational modes.

$$H_B = \frac{1}{4} \sum_r \left( \hbar \omega_r B_r^+ B_r + \hbar \tilde{\omega}_r \tilde{B}_r^+ \tilde{B}_r \right) \tag{3.6}$$

The essential problem of the Born-Oppenheimer approximation lies in the fact, that initially the electronic states are quantized whereas the motion of nuclei remains in classical form. Then the transition from the Cartesian to the normal coordinates is carried out on the basis of Newton mechanics, and finally the nuclear motion is quantized as the system of independent harmonic oscillators. This procedure represents the hierarchical type of quantization, which is a complete contradiction of the fundamental requirement of the second quantization procedure of the total Hamiltonian that must be simultaneous.

If we would try to find an ontological interpretation of the Born-Oppenheimer hierarchical type of quantization, surely the concept of quantization of atomic centers would be more adequate description than the concept of quantization of nuclei motion. The question arises how to proceed back to the simultaneous quantization of the unit system of electrons and nuclei. I have solved this problem in my PhD thesis for the quantization of electron-vibrational Hamiltonian. Now I present the revised version for the complete electron-hypervibrational one because the original version has totally failed during the comparison tests with the Born-Handy ansatz results.

We cannot start with the B-O approximation of moving electrons, following the motion of nuclei, but with the crude representation, i.e. the representation of fixed nuclear positions. General form of nonrelativistic Hamiltonian for any molecular and crystal system can be written in the form

$$H = T_N(\bar{\bar{B}}) + E_{NN}(\bar{B}) + \sum_{PQ} h_{PQ}(\bar{B})\bar{a}_P^+\bar{a}_Q + \frac{1}{2}\sum_{PQRS} v_{PQRS}^0 \bar{a}_P^+\bar{a}_Q^+\bar{a}_S\bar{a}_R \qquad (3.7)$$

where $T_N$ stands for kinetic energy of nuclei and $E_{NN}$ for the potential energy of nuclei interactions. One-electron matrix elements $h_{PQ}$ comprise electron kinetic energy and electron-nuclear interaction. The term $v_{PQRS}^0$ represents two-electron interaction matrix elements. The operators marked with the bar are operators of "original" quasiparticles (electrons and hyperphonons) in the crude representation. The terms $E_{NN}$ and $h_{PQ}$ are defined through their Taylor expansion (see equations (A1-4) in Appendix A for details)

$$E_{NN}(\bar{B}) = \sum_{n=0}^{\infty} E_{NN}^{(n)}(\bar{B}) \qquad (3.8)$$

$$h_{PQ}(\bar{B}) = h_{PQ}^0 + \sum_{n=1}^{\infty} u_{PQ}^{(n)}(\bar{B}) \qquad (3.9)$$

where $h_{PQ}^0$ is one-electron term for fixed (equilibrium) nuclear coordinates and

$$u_{PQ}(\bar{B}) = \langle P | \sum_i \frac{-Z_i e^2}{|\mathbf{r} - \mathbf{R}_i|} | Q \rangle \qquad (3.10)$$

in terms of the second quantization represents the matrix elements of electron-hypervibrational interaction. We assume in (3.8,9) that the sums are convergent. It is important to emphasize that the operators $\bar{B}$ and $\bar{\bar{B}}$ in (3.7) refer to the whole set of hypervibrations.

Potential energy of nuclei motion is defined through the quadratic part of internuclear potential plus some additive term representing the selfconsistent influence of electron-nuclear potential

$$E_{pot} = E_{NN}^{(2)}(\bar{B}) + V_N^{(2)}(\bar{B}) \qquad (3.11)$$

In the adiabatic limit the values of $V_N^{(2)}$ can be evaluated simply through the coupled perturbed Hartree-Fock method and the kinetic energy $E_{kin}$ is identical with the kinetic energy of nuclei $T_N$. Now the crucial step is coming: At the case when the adiabatic approximation is not valid it is necessary to incorporate the new additive kinetic term originating from the kinetic energy of electrons. The resulting kinetic energy of the system has the form

$$E_{kin} = T_N(\bar{\bar{B}}) + W_N^{(2)}(\bar{\bar{B}}) \qquad (3.12)$$

The total Hamiltonian (3.7) can be now divided into two parts

$$H = H_A + H_B \qquad (3.13)$$

Where the first part $H_A$ reads

$$H_A = E_{NN}(\overline{B}) - E_{NN}^{(2)}(\overline{B}) - V_N^{(2)}(\overline{B}) - W_N^{(2)}(\overline{\overline{B}}) + \sum_{PQ} h_{PQ}(\overline{B}) \overline{a}_P^+ \overline{a}_Q$$

$$+ \frac{1}{2} \sum_{PQRS} v_{PQRS}^0 \overline{a}_P^+ \overline{a}_Q^+ \overline{a}_S \overline{a}_R \tag{3.14}$$

and the second part $H_B$ has the same form as (3.4)

$$H_B = \frac{1}{4} \sum_r \left( \hbar \omega_r \overline{B}_r^+ \overline{B}_r + \hbar \tilde{\omega}_r \overline{\overline{B}}_r^+ \overline{\overline{B}}_r \right) \tag{3.15}$$

The final electron-hypervibrational Hamiltonian in the second quantization formalism has now the form

$$H = E_{NN}(\overline{B}) - E_{NN}^{(2)}(\overline{B}) - V_N^{(2)}(\overline{B}) - W_N^{(2)}(\overline{\overline{B}}) + \sum_{PQ} h_{PQ}(\overline{B}) \overline{a}_P^+ \overline{a}_Q$$

$$+ \frac{1}{2} \sum_{PQRS} v_{PQRS}^0 \overline{a}_P^+ \overline{a}_Q^+ \overline{a}_S \overline{a}_R + \frac{1}{4} \sum_r \left( \hbar \omega_r \overline{B}_r^+ \overline{B}_r + \hbar \tilde{\omega}_r \overline{\overline{B}}_r^+ \overline{\overline{B}}_r \right) \tag{3.16}$$

It is necessary to notice that the crude representation (3.16) is the first and the last one where the quantization of nuclear motion can be accomplished by means of the classical Newton mechanics separation of degrees of freedom. All other representations will mix the vibrational, rotational and translational modes, and they will be not separable any more.

# 4. Unitary transformations applied to the electron-hypervibrational Hamiltonian

Our aim is now to find the most general group of quasiparticle transformations for electron fermion and hypervibration boson operators, binding individual representations of the total Hamiltonian. I have proposed in my first work on this topic [12] two transformations - the first of the adiabatic type, dependent on nuclear coordinates $Q$, and the second of nonadiabatic type, dependent on the nuclear momenta $P$. When I tried later to apply the results to solid state physics, I have recognized that my transformations, applied on individual fermion and boson operators, are identical with the well-known Fröhlich transformation [10], applied on the whole Hamiltonian

$$H' = e^{-S(Q,P)} H e^{S(Q,P)} \tag{4.1}$$

namely with its decomposed part

$$H' = e^{-S_2(P)} e^{-S_1(Q)} H e^{S_1(Q)} e^{S_2(P)} \tag{4.2}$$

Although both expressions (4.1) and (4.2) are equivalent, they differ a little bit in individual orders of the Taylor expansion since the coordinate and momentum operators do not commute. This difference is not significant but yet the latter one is more convenient due to symmetric properties of final expressions as it will be discussed hereafter.

Fröhlich transformation is essential. It was applied to superconductivity theory but has not been used in quantum chemistry problems, however.

But unfortunately Fröhlich could not know that electron-phonon interaction is not the true interaction archetype and applied his transformation to a wrong Hamiltonian. I did not initially involved the COM problem into consideration, and so I have made the same mistake as Fröhlich in all previous works [12-20] with the negative consequences for the later comparison tests [22] with the Born-Handy ansatz. Therefore I present now a solution with incorporation of all 3N degrees of freedom, unified in the concept of electron-hyperphonon interaction.

The advantage of the quasiparticle transformations lies in the fact that they are more transparent than the global transformation of the whole Hamiltonian. The first transformation in (4.2) with generator $S_1$ is equivalent to the adiabatic quasiparticle transformation from the crude into the adiabatic representation, defined through new quasiparticles in adiabatic representation with double bar

$$\bar{a}_P = \sum_Q c_{PQ}(\bar{\bar{B}})\bar{\bar{a}}_Q \tag{4.3}$$

$$\bar{b}_r = \bar{\bar{b}}_r + \sum_{PQ} d_{rPQ}(\bar{\bar{B}})\bar{\bar{a}}_P^+\bar{\bar{a}}_Q \tag{4.4}$$

Analogical equations hold for the creation operators. The operators $c_{PQ}(\bar{\bar{B}})$ and $d_{rPQ}(\bar{\bar{B}})$ are defined trough their Taylor expansions and are limited through the unitary conditions

$$\sum_R c_{PR}c_{QR}^+ = \delta_{PQ} \tag{4.5}$$

$$d_{rPQ} = \sum_R c_{RP}^+[\bar{\bar{b}}_r, c_{RQ}] \tag{4.6}$$

The second transformation with generator $S_2$ is equivalent to the nonadiabatic transformation from the adiabatic representation into the final one, which we shall call „general", i.e. the representation that involves the adiabatic case as well as the nonadiabatic one. This representation is defined through new quasiparticles denoted simply without bar

$$\bar{\bar{a}}_P = \sum_Q \tilde{c}_{PQ}(\tilde{B})a_Q \tag{4.7}$$

$$\bar{\bar{b}}_r = b_r + \sum_{PQ} \tilde{d}_{rPQ}(\tilde{B})a_P^+ a_Q \tag{4.8}$$

where the operators $\tilde{c}_{PQ}(\tilde{B})$ and $\tilde{d}_{rPQ}(\tilde{B})$ are defined through their Taylor expansions and are limited through the unitary conditions

$$\sum_R \tilde{c}_{PR}\tilde{c}_{QR}^+ = \delta_{PQ} \tag{4.9}$$

$$\tilde{d}_{rPQ} = \sum_R \tilde{c}_{RP}^+[b_r, \tilde{c}_{RQ}] \tag{4.10}$$

For the Taylor expansion of both adiabatic and nonadiabatic unitarity conditions see equations (A5-8) in Appendix A.

The form of the transformed Hamiltonian is very complex and the individual terms are put into Appendix B. We demonstrate now only the main steps of treatment of the transformed Hamiltonian in the general representation.

At the first stage we will apply the Wick's theorem, as it is standardly defined in quantum chemistry, i.e. with respect to Fermi vacuum. For one-fermion terms the Wick's theorem results in

$$\sum_{PQ} \lambda_{PQ} a_P^+ a_Q = \sum_{PQ} \lambda_{PQ} N[a_P^+ a_Q] + \sum_I \lambda_{II} \tag{4.11}$$

and for two-fermion terms

$$\sum_{PQRS} \mu_{PQRS} a_P^+ a_Q^+ a_S a_R = \sum_{PQRS} \mu_{PQRS} N[a_P^+ a_Q^+ a_S a_R]$$
$$+ \sum_{PQI} (\mu_{PIQI} + \mu_{IPIQ} - \mu_{PIIQ} - \mu_{IPQI}) N[a_P^+ a_Q] + \sum_{IJ} (\mu_{IJIJ} - \mu_{IJJI}) \tag{4.12}$$

Analogical relations hold for three-fermion terms, which also occur in the transformed Hamiltonian. After complex application of the Wick's theorem on all fermion operators we get the normal form of the Hamiltonian in the general representation.

At the second stage we perform the very well known Moller-Plesset splitting [28] of the final Hamiltonian, i.e. the diagonalization of one-fermion terms in the normal form according to the formula

$$\sum_{PQ} \lambda_{PQ} N[a_P^+ a_Q] \rightarrow \sum_P \Lambda_P N[a_P^+ a_P] \tag{4.13}$$

First we obtain the well-known Hartree-Fock equation

$$f_{PQ}^0 = h_{PQ}^0 + \sum_I (v_{PQII}^0 - v_{PIIQ}^0) = \varepsilon_P^0 \delta_{PQ} \tag{4.14}$$

Diagonalization of the terms which contain boson operators in the first order gives us equations for the first order coefficients of the unknown operators $c$ and $\tilde{c}$ of quasiparticle transformations (4.3) and (4.7)

$$u_{PQ}^r + (\varepsilon_P^0 - \varepsilon_Q^0) c_{PQ}^r + \sum_{AI} [(v_{PIQA}^0 - v_{PIAQ}^0) c_{AI}^r - (v_{PAQI}^0 - v_{PAIQ}^0) c_{IA}^r] - \hbar \omega_r \tilde{c}_{PQ}^r = \varepsilon_P^r \delta_{PQ}$$
$$c_{PP}^r = 0 \tag{4.15}$$
$$(\varepsilon_P^0 - \varepsilon_Q^0) \tilde{c}_{PQ}^r + \sum_{AI} [(v_{PIQA}^0 - v_{PIAQ}^0) \tilde{c}_{AI}^r - (v_{PAQI}^0 - v_{PAIQ}^0) \tilde{c}_{IA}^r] - \hbar \tilde{\omega}_r c_{PQ}^r = \tilde{\varepsilon}_P^r \delta_{PQ}$$
$$\tilde{c}_{PP}^r = 0 \tag{4.16}$$

with the simplest chosen calibration for the diagonal terms.

Finally the set of equations in the second order of the Taylor expansion results in the ab-initio selfconsistent equations for hypervibrational frequencies $\omega$ and $\tilde{\omega}$, namely for unknown potential and kinetic matrix elements (3.11) and (3.12)

$$V_N^{rs} = \sum_I u_{II}^{rs} + \sum_{AI} [(u_{IA}^r + \hbar \omega_r \tilde{c}_{IA}^r) c_{AI}^s + (u_{IA}^s + \hbar \omega_s \tilde{c}_{IA}^s) c_{AI}^r] \tag{4.17}$$

$$W_N^{rs} = 2\hbar \tilde{\omega}_r \sum_{AI} c_{AI}^r \tilde{c}_{IA}^s \tag{4.18}$$

We can look at equations (4.15-18) as the generalization of the CPHF [23,24] equations for the case of general representation, i.e. including the cases of break-down of the B-O approximation. We shall call them COM CPHF equations.

In the adiabatic limit where the coefficients $\tilde{c}$ equal zero we get

$$u_{PQ}^r + (\varepsilon_P^0 - \varepsilon_Q^0)c_{PQ}^r + \sum_{AI}[(v_{PIQA}^0 - v_{PIAQ}^0)c_{AI}^r - (v_{PAQI}^0 - v_{PAIQ}^0)c_{IA}^r] = \varepsilon_P^r \delta_{PQ}; \quad c_{PP}^r = 0 \tag{4.19}$$

$$V_N^{rs} = \sum_I u_{II}^{rs} + \sum_{AI}\left(u_{IA}^r c_{AI}^s + u_{IA}^s c_{AI}^r\right) \tag{4.20}$$

Adiabatic limit supposes that the vibrational frequencies $\omega$ are much smaller than the difference $\Delta\varepsilon^0$ between the first unoccupied and last occupied orbital. If we estimate the ratio $\tilde{c}/c$ from equations (4.15-16) we can find out that the proportion holds

$$\tilde{c} \sim c\frac{\omega}{\Delta\varepsilon^0} \tag{4.21}$$

It means that the accuracy of adiabatic equations (4.19-20) is limited up to the first order of the ratio $\omega/\Delta\varepsilon^0$. Adiabatic representation still binds vibrations, rotations and translations into one set of nonseparable hypervibrations.

Now we proceed from the adiabatic limit to the B-O limit. In both approximations the same equations (4.19-20) hold. The only but remarkable difference is the classical concept of the separation of degrees of freedom in the latter one. It means that the coefficients $r,s$ in these equations represent only the normal vibrational modes. And just in this simplified form the equations (4.19-20) are exactly identical with the standard Pople's CPHF equations [23,24] after the formal rewrite from the fixed basis of atomic orbitals into the moving one, following the motion of nuclei. Since this is only the numerical problem, which does not affect the substance of this topic, we only refer to preceding works [12,17].

## 5. Derivation of the extended Born-Handy ansatz from the general representation

Let us proceed to the fermion part of the general Hamiltonian, particularly the fermion part difference $\Delta H_F$ between the general Hamiltonian and the original crude one

$$H_F = H_{F(0)} + \Delta H_F \tag{5.1}$$

The Hamiltonian $H_{F(0)}$ consists of three well-known parts

$$H_{F(0)} = H_{F(0)}^0 + H_{F(0)}' + H_{F(0)}'' \tag{5.2}$$

where

$$H_{F(0)}^0 = E_0 = E_{NN}^0 + E_{SCF}^0 = E_{NN}^0 + \sum_I h_{II}^0 + \frac{1}{2}\sum_{IJ}(v_{IIJJ}^0 - v_{IJJI}^0) \tag{5.3}$$

contains the SCF energy of unperturbed electronic system and the nuclear potential energy,

$$H_{F(0)}' = \sum_P \varepsilon_P^0 N[a_P^+ a_P] \tag{5.4}$$

is one-electron spectrum as a result of diagonalization (4.14) and

$$H_{F(0)}'' = \frac{1}{2}\sum_{PQRS} v_{PQRS}^0 N[a_P^+ a_Q^+ a_S a_R] \tag{5.5}$$

represents two-electron Coulomb interaction in a normal product form.

The most interesting is the Hamiltonian $\Delta H_F$ consisting of four parts. As one can see in Appendix C the transformations produce also the three-fermion terms. Because they are irrelevant for further considerations we limit the study only to the three important parts:

$$\Delta H_F = \Delta H_F^0 + \Delta H_F' + \Delta H_F''$$  (5.6)

For the correction to the ground state energy we get

$$\Delta H_F^0 = \Delta E_0 = \sum_{AIr} \left( \hbar \tilde{\omega}_r \mid c_{AI}^r \mid^2 - \hbar \omega_r \mid \tilde{c}_{AI}^r \mid^2 \right)$$  (5.7)

The one-particle correction $\Delta H_F'$ is more complex and therefore we select only that terms which are decisive for excitation mechanism

$$\Delta H_F' = \sum_{PQr} \left[ \hbar \tilde{\omega}_r \left( \sum_A c_{PA}^r c_{QA}^{r*} - \sum_I c_{PI}^r c_{QI}^{r*} \right) - \hbar \omega_r \left( \sum_A \tilde{c}_{PA}^r \tilde{c}_{QA}^{r*} - \sum_I \tilde{c}_{PI}^r \tilde{c}_{QI}^{r*} \right) \right] N[a_P^+ a_Q]$$
$$+ \sum_{PRr} \left[ \left( \varepsilon_P^0 - \varepsilon_R^0 \right) \left( \mid c_{PR}^r \mid^2 + \mid \tilde{c}_{PR}^r \mid^2 \right) - 2 \hbar \tilde{\omega}_r \, \mathrm{Re} \left( \tilde{c}_{PR}^r c_{PR}^{r*} \right) \right] N[a_P^+ a_P]$$  (5.8)

The first part (5.8) is of a pure one-fermion origin and in the complete derivation (see Appendix C) has a non-diagonal form. The second part is not of a pure one-fermion origin. It is a vacuum value of type $\langle 0 \mid B_r B_s \mid 0 \rangle$ and/or $\langle 0 \mid \tilde{B}_r \tilde{B}_s \mid 0 \rangle$ of the mixed fermion-boson terms, where the bosonic part is of the quadratic form of coordinate and/or momentum operators.

In a similar way we select from the correction $\Delta H_F''$ only the dominant term

$$\Delta H_F'' = \sum_{PQRSr} \left( \hbar \tilde{\omega}_r c_{PR}^r c_{SQ}^{r*} - \hbar \omega_r \tilde{c}_{PR}^r \tilde{c}_{SQ}^{r*} \right) N[a_P^+ a_Q^+ a_S a_R]$$  (5.9)

If we proceed from the general to the adiabatic representation with zero $\tilde{c}$ coefficients we obtain exactly the Born-Handy ansatz (2.14) from the first principle derivation:

$$\Delta E_{0(ad)} = \sum_{AIr} \hbar \tilde{\omega}_r \mid c_{AI}^r \mid^2$$  (5.10)

Moreover, this expression is fully covariant with respect to all degrees of freedom, i.e. vibrations, rotations and translations.

Whereas the Born-Handy formula holds only in the framework of adiabatic approximation, the equation (5.7) is quite general and holds in the whole scale $\omega / \Delta \varepsilon^0$ including the non-adiabatic cases where the B-O approximation is broken. We can call it the generalized or extended Born-Handy formula. It is of enormous importance because, as we will see further, it defines the molecular and crystallic structure.

For a deeper insight into the properties of the extended Born-Handy formula we demonstrate its simple solution neglecting the two-electron terms. Then the equations (4.15-16) have the analytical solution for the coefficients $c$ and $\tilde{c}$ :

$$c_{PQ}^r = u_{PQ}^r \frac{\varepsilon_P^0 - \varepsilon_Q^0}{(\hbar \omega_r)^2 - (\varepsilon_P^0 - \varepsilon_Q^0)^2}$$  (5.11)

$$\tilde{c}_{PQ}^r = u_{PQ}^r \frac{\hbar \tilde{\omega}_r}{(\hbar \omega_r)^2 - (\varepsilon_P^0 - \varepsilon_Q^0)^2}$$  (5.12)

so that the extended Born-Handy formula can be expressed only by means of the matrix elements of electron-hypervibrational (electron-hyperphonon) interaction, one-electron energies and hypervibrational (hyperphonon) frequencies.

$$\Delta E_0 = \sum_{AIr} |u_{AI}^r|^2 \frac{\hbar \tilde{\omega}_r}{(\varepsilon_A^0 - \varepsilon_I^0)^2 - (\hbar \omega_r)^2} \tag{5.13}$$

Rewriting this equation in the form of the sum of vibrational, rotational and translational parts, we obtain

$$\Delta E_0 = \sum_{AI, r \in V} |u_{AI}^r|^2 \frac{\hbar \omega_r}{(\varepsilon_A^0 - \varepsilon_I^0)^2 - (\hbar \omega_r)^2}$$

$$+ 2 \sum_{AI, r \in R} |u_{AI}^r|^2 \frac{\rho_r}{(\varepsilon_A^0 - \varepsilon_I^0)^2} + 2 \sum_{AI, r \in T} |u_{AI}^r|^2 \frac{\tau_r}{(\varepsilon_A^0 - \varepsilon_I^0)^2} \tag{5.14}$$

and so we have separate expressions for electron-vibrational (electron-phonon), electron-rotational (electron-roton) and electron-translational (electron-translon) contributions of the extended Born-Handy formula. In a full analogy with phonons – quasiparticles generated by the nuclear vibrations – we introduce the similar quasiparticles for the nuclear rotations and translations and we will call them simply rotons and translons.

In the case of adiabatic approximation where the inequality $\hbar \omega_r \ll \varepsilon_A^0 - \varepsilon_I^0$ holds the equation (5.14) represents only a small correction to the energy of groundstate. But what about the non-adiabatic case? In the electron-phonon part of (5.14) there is a possible singularity when $\hbar \omega_r \approx \varepsilon_A^0 - \varepsilon_I^0$. This case should affect the arising singularities in equations (5.11-12) for the coefficients $c$ and $\tilde{c}$. Fortunately this theory is fully selfconsistent and the extreme values of these coefficients should affect backward the frequencies $\omega_r$ defined through the equations (4.17-18) so that the system has its own self-defense against such type of singularities, at least in molecules where the number of vibrational modes is finite. In crystals with infinite number of vibrational modes the integration valuer principal is used which eliminates this type of singularities.

The last two parts – electron-roton and electron-translon – of the equation (5.14) have singularities in the case of groundstate electron degeneracies. There are only two ways how to avoid them. Either all matrix elements $u_{AI}^r$ for the degenerate states $A$ and $I$ and for rotational and translational modes have to equal zero, or the whole system has to change its structural arrangement in order to remove the degeneracy. The new equilibrium position of the nuclei naturally somehow increase the selfconsistent energy given by equation (5.3) but on the other hand the final energy of the ground state can be still smaller due to the first part of (5.14) which is negative in the case when both inequalities hold: $\varepsilon_A^0 - \varepsilon_I^0 > 0$ and $\varepsilon_A^0 - \varepsilon_I^0 < \hbar \omega_r$.

Now we can summarize the previous considerations: In the case of the break down of B-O approximation the electron-roton and electron-translon parts of the extended Born-Handy formula play the role of the trigger inducing a structural instability in the system. These parts are direct consequence of the introduction of centre-of-mass problem into account. They are responsible for the formation of molecular and crystallic structure. On the other hand, the electron-phonon part plays the role of stabilizer in a new equilibrium position with new nuclear displacements. Therefore it is responsible for the formation of molecular and crystallic electronic structure.

## 6. Jahn-Teller effect

Jahn-Teller (J-T) effect is a direct consequence of the breakdown of the B-O approximation. For the first time this effect was studied only in a qualitative way on the basis of the group theory [7]. Nowadays there exist many extensive monographies dealing with the exact solutions of simple models where two degenerate or quasidegenerate levels are usually coupled with one or two vibrational modes [29,30].

J-T effect deals with the molecular distortions due to electronically degenerate ground states. The J-T theorem was formulated as a statement: "For non-linear molecular entities in a geometry described by a point symmetry group possessing degenerate irreducible representations there always exists at least one non-totally symmetric vibration that makes electronically degenerate states unstable at this geometry. The nuclei are displaced to new equilibrium positions of lower symmetry causing a splitting of the originally degenerate states."

There were only few articles devoted to the J-T effect in the thirties before the World War II. Then the period of stagnation lasted almost two decades. Bersuker in his book [8] describes the reason:

Among other things Van Vleck [31] wrote that "it is a great merit of the J-T effect that it disappears when not needed." This declaration reflects the situation when there was very poor understanding of what observable effects should be expected as a consequence of the J-T theorem. The point is that the simplified formulation of the consequences of the J-T theorem as "spontaneous distortion" is incomplete and therefore inaccurate, and may lead to misunderstanding. In fact, there are several (or an infinite number of) equivalent directions of distortion, and the system may resonate between them (the dynamic J-T effect)... It does not necessarily lead to observable nuclear configuration distortion, and this explains why such distortions often cannot be observed directly... Even in 1960 Low in his book [32] stated that "it is a property of the J-T effect that whenever one tries to find it, it eludes measurements."

The theoretical methods of calculation of the J-T effect started to be developed in the fifties, after the first experimental confirmations appeared. These methods are based on perturbation theory, in which the influence of the nuclear displacements via electron–vibrational (vibronic) interactions is considered as a perturbation to the degenerate states, and moreover, they are considered to be the proof of the J-T theorem.

Let us focus on the origin of the principal idea of the J-T effect. Before its final formulation by Jahn and Teller, first the Teller's student Renner [33] was inspired with the von Neumann–Wigner theorem about crossing electronic terms [34]: "Electronic states of a diatomic molecule do not cross, unless permitted by symmetry". Only if the states have different symmetry, they can cross.

When we look more carefully on the von Neumann-Wigner and J-T theorems we can see the significant difference: Whereas the first one is connected with the dissociation processes in molecules and the question of crossing or non-crossing their potential curves, the second one concerns the rigid molecules and the question of their equilibrium nuclear positions. J-T effect was formulated prematurely without the exact knowledge what forms the molecular structure and what the break down of B-O approximation really means. Two important factors were never incorporated in the J-T effect: the Fröhlich transformation and the centre-of-mass problem. These two factors are so emergent that the J-T effect will never be explained correctly without them.

After the inclusion of the COM separation into the considerations we can immediately recognize from the equation (5.14), that vibronic coupling is not responsible for the J-T effect at all. The authentic J-T trigger is represented by the electron-translational (translonic) and electron-rotational (rotonic) coupling.

Nowadays there are many attempts to implement the J-T effect into the superconductivity problem. But first something related to superconductivity has to be implemented into the J-T effect: the Fröhlich transformation. Fröhlich has proposed his transformation [10] almost twenty years after the first formulation of the J-T effect [33]. Unfortunately, this transformation is known only in solid state physics (and moreover used in practice solely in the superconductivity problem) but after more than a half of century it was never integrated in quantum chemistry. It is very important for the explanation of the hypervibronic coupling mechanism in the J-T effect. It takes into account not only the dependence of electronic states on the nuclear coordinates, as it is usual in the adiabatic case, but also on the nuclear momenta, which is inherent in the non-adiabatic one. This type of transformation leads to new fermion quasiparticles that are able to describe the non-adiabatic case on the one-determinantal level, so that no secular problem with the crossing degenerate states arises as it is standardly used in the J-T calculations.

As we can see from the equation (5.14), after the translonic and rotonic coupling evokes the structural change, and therefore the small changes of the unperturbed energies $\varepsilon^0$, too, the vibronic coupling stabilizes the system in this new position. Let us look at the one-partical corrections $\Delta\varepsilon$ to these new unperturbed energies $\varepsilon^0$. For illustration we show only the diagonal corrections from the equation (5.8) in an analytical form after the neglecting of two-electron terms. Substituting from equations (5.11-12) into (5.8) we get:

$$\Delta H_F' = \sum_P \Delta \varepsilon_P N[a_P^+ a_P]$$

$$= \sum_{Pr} \hbar \tilde{\omega}_r \left( \sum_{A \neq P} \frac{|u_{PA}^r|^2}{(\varepsilon_P^0 - \varepsilon_A^0)^2 - (\hbar \omega_r)^2} - \sum_{I \neq P} \frac{|u_{PI}^r|^2}{(\varepsilon_P^0 - \varepsilon_I^0)^2 - (\hbar \omega_r)^2} \right) N[a_P^+ a_P] \qquad (6.1)$$

In order to have a better comparison of equations for one-electron energies (5.8), (6.1) and those for corrections to the ground state (5.7), (5.13) we introduce a symmetrical matrix $\boldsymbol{\Omega}$ [13,19].

$$\Omega_{PQ} = \sum_r \left( \hbar \tilde{\omega}_r \mid c_{PQ}^r \mid^2 - \hbar \omega_r \mid \tilde{c}_{PQ}^r \mid^2 \right) = \sum_r |u_{PQ}^r|^2 \frac{\hbar \tilde{\omega}_r}{(\varepsilon_P^0 - \varepsilon_Q^0)^2 - (\hbar \omega_r)^2}$$

$$\Omega_{PQ} = \Omega_{QP}; \qquad \Omega_{PP} = 0 \qquad (6.2)$$

After the substitution of (6.2) into the aforementioned equations we get

$$\Delta E_0 = \sum_{AI} \Omega_{AI} \qquad (6.3)$$

$$\Delta \varepsilon_P = \sum_A \Omega_{PA} - \sum_I \Omega_{PI} \qquad (6.4)$$

Let us demonstrate the solution of the extended Born-Handy formula on an example of J-T effect with two degenerate electronic states. First rotonic and translonic coupling split them, so we obtain in a closed shell case one occupied orbital and one unoccupied (virtual) orbital with the unperturbed energies $\varepsilon_o^0$ and $\varepsilon_u^0$. Then the system finds its new equilibrium position via the vibronic coupling with the minimal value of the total energy. Therefore we have

$$\Delta E_0 = 2 \Omega_{uo} < 0 \qquad (6.5)$$

and consequently the following relations hold

$$\Delta \varepsilon_o = \Omega_{uo} < 0; \qquad \Delta \varepsilon_u = -\Omega_{uo} > 0 \qquad (6.6)$$

We have proved in this way the J-T one-electron energy splitting as a direct consequence of the solution of the extended Born-Handy formula.

The above mentioned considerations have a very important consequences. They imply that the extended Born-Handy formula, derived from the first principles, is the factual master equation for explanation as well as for calculation of the J-T effect. This effect was discovered too soon without the proper quantum mechanical knowledge. The inspiration in the von Neumann-Wigner theorem was out of the depth of this problem, it could result at most in the group theory formulation of the J-T effect. Because the Fröhlich transformation, respecting the influence of nuclear momenta on electronic states, and the center-of-mass problem, leading to the critical solution in the case of degenerate electronic states, were not taken into account, the J-T effect was falsely justified by the way of vibronic coupling. The authentic trigger of the J-T effect is not the vibronic, but rotonic and translonic coupling.

It is therefore necessary to reformulate the J-T effect, and not only in a version for molecules, but for both – molecules and crystals. The ontological statement emanating from the extended Born-Handy formula (5.14) is essential; all other considerations regarding the symmetrical properties of molecules and crystal, and of electronic states and vibration - rotation - translation modes follow as a consequence of the properties of this formula. Here is the draft of the reformulated J-T theorem:

**Molecular and crystallic entities in a geometry of electronically degenerate ground state are unstable at this geometry except the case when all matrix elements of electron-rotational and electron-translational interaction equal zero.**

## 7. Conductivity

Let us focus now in the case when all matrix elements of electron-rotational and electron-translational interaction equal zero. Then the system geometry of electronically degenerate ground state survives. This is exactly the case of conductors in solid state physics. The electron-hypervibrational problem reduces to a simple classical electron-vibrational one. Equation (5.14) has then the form:

$$\Delta E_0 = \sum_{Al,r \in V} |u_{Al}^r|^2 \frac{\hbar \omega_r}{(\varepsilon_A^0 - \varepsilon_I^0)^2 - (\hbar \omega_r)^2} \tag{7.1}$$

The question arises, how to achieve the nullification of all electron-rotational and electron-translational terms $u_{Al}^r$ in (5.14). If we try to find the solution in the form of the Bloch functions (which fully reflect the symmetry of the crystal), then since rotons and translons have zero quasimomentum values and virtual and occupied states $A$ and $I$ correspond to different quasimomentum values $\mathbf{k}$ and $\mathbf{k'}$, naturally the above mentioned requirement for $u_{Al}^r$ is fulfilled.

Now we can rewrite the equation (7.1) in solid state notation ($r \rightarrow \mathbf{q}$; $I \rightarrow \mathbf{k}, \sigma$ with the occupation factor $f_{\mathbf{k}}$; $A \rightarrow \mathbf{k'}, \sigma'$ with the occupation factor $1-f_{\mathbf{k'}}$; $\varepsilon_I^0 \rightarrow \varepsilon_{\mathbf{k}}^0$; $\varepsilon_A^0 \rightarrow \varepsilon_{\mathbf{k'}}^0$; $u_{Al}^r \rightarrow u_{\mathbf{k'k}}^{\mathbf{q}} = u^{\mathbf{k'-k}} = u^{\mathbf{q}}$).

$$\Delta E_0 = 2 \sum_{\mathbf{k}, \mathbf{k'}; \mathbf{k} \neq \mathbf{k'}} |u^{\mathbf{k'-k}}|^2 f_{\mathbf{k}}(1-f_{\mathbf{k'}}) \frac{\hbar \omega_{\mathbf{k'-k}}}{(\varepsilon_{\mathbf{k'}}^0 - \varepsilon_{\mathbf{k}}^0)^2 - (\hbar \omega_{\mathbf{k'-k}})^2} \tag{7.2}$$

This formula was derived by Fröhlich by means of the second order perturbation theory [9] and rederived by means of the unitary transformation [10]. Fröhlich in his derivations started with the generally accepted Hamiltonian in solid state physics:

$$H = \sum_{\mathbf{k}, \sigma} \varepsilon_{\mathbf{k}} a_{\mathbf{k}, \sigma}^+ a_{\mathbf{k}, \sigma} + \sum_{\mathbf{q}} \hbar \omega_{\mathbf{q}} \left( b_{\mathbf{q}}^+ b_{\mathbf{q}} + \tfrac{1}{2} \right) + \sum_{\mathbf{k}, \mathbf{q}, \sigma} u^{\mathbf{q}} \left( b_{\mathbf{q}} + b_{-\mathbf{q}}^+ \right) a_{\mathbf{k+q}, \sigma}^+ a_{\mathbf{k}, \sigma} \tag{7.3}$$

And just the equation (7.3) is the crucial problem. It involves only the electron-phonon terms and not the electron-hyperphonon ones which are necessary for the explanation of superconductors. This equation can be a good starting point for insulators, semiconductors and conductors, but never for superconductors. Fröhlich first believed that through the optimization of occupation factors $f_{\mathbf{k}}$ in (7.2) he gets some decrease of the total energy and tried to interpret this new state as the state of superconductors, but later recognized that this solution leads to no experimentally detected gap. Although his transformations were unique and marvellous, they were unfortunately applied to the absolutely wrong Hamiltonian. Without knowing it, Fröhlich derived in his equation (7.2) exactly the correlation energy to the ground state of conductors.

After bypassing the trigger in (5.14), on the one hand the crystal remains in the adiabatic state, but on the other hand the electrons from the last occupied (conducting) band are not the part of the rigid system any more, they are quasi free and interact with the lattice only via the electron-phonon interaction without the backward influence on the lattice symmetry and nuclear displacements. The whole system is divided in two subsystems, the adiabatic „core" consisting of nuclei and electron valence bands, and the quasi free conducting electrons.

Therefore the equation (7.3) describes the crude representation with the energies of these two subsystems and the interaction terms between them. Justification of its use for conductors is not given a priori but as a consequence of the abnormal solution of the extended Born-Handy formula which bypasses the trigger initiating the J-T effect, so that the general electron-hyperphonon problem can reduce to the simple electron-phonon one. The explanation of conductivity is not at all so simple as it is universal believed. The COM problem plays here an important role and the conductivity represents only one possible solution of this problem. Fortunately, since this solution of the COM problem fully justifies the validity of the Hamiltonian (7.3) for conductors, all equations derived for them remain correct even though the COM problem was not included into the consideration.

Let us proceed to the one-electron corrections (5.8) when the electron-hypervibrational problem reduces to the electron-vibrational one. Since conductors have no gap, the one-particle derivation is more sensitive and we have to take into account also the second part of (5.8) which does not depend on the electron distribution defined by occupation factors as the first part but is exactly valid in the present form only for the vibrational vacuum.

The substitution for $c$ and $\tilde{c}$ in (5.11-12) gives us

$$\Delta H_F' = \sum_{P,r \in V} \left( \sum_{A \neq P} \frac{|u_{PA}^r|^2}{\varepsilon_P^0 - \varepsilon_A^0 - \hbar\omega_r} + \sum_{I \neq P} \frac{|u_{PI}^r|^2}{\varepsilon_P^0 - \varepsilon_I^0 + \hbar\omega_r} \right) N[a_P^+ a_P]$$

$$= \sum_{P,r \in V} \left( \sum_{R \neq P} |u_{PR}^r|^2 \frac{1}{\varepsilon_P^0 - \varepsilon_R^0 - \hbar\omega_r} - 2 \sum_{I \neq P} |u_{PI}^r|^2 \frac{\hbar\omega_r}{(\varepsilon_A^0 - \varepsilon_I^0)^2 - (\hbar\omega_r)^2} \right) N[a_P^+ a_P] \tag{7.4}$$

and in the solid state notation ($r \rightarrow \mathbf{q}$; $P \rightarrow \mathbf{k},\sigma$; $R \rightarrow \mathbf{k-q},\sigma$; $I \rightarrow \mathbf{k-q},\sigma$ with the occupation factor $f_{\mathbf{k-q}}$)

$$\Delta H_F' = \sum_{\mathbf{k},\mathbf{q},\sigma;\,\mathbf{q} \neq 0} |u^{\mathbf{q}}|^2 \frac{1}{\varepsilon_{\mathbf{k}}^0 - \varepsilon_{\mathbf{k-q}}^0 - \hbar\omega_{\mathbf{q}}} N[a_{\mathbf{k},\sigma}^+ a_{\mathbf{k},\sigma}]$$

$$- 2 \sum_{\mathbf{k},\mathbf{q},\sigma;\,\mathbf{q} \neq 0} |u^{\mathbf{q}}|^2 f_{\mathbf{k-q}} \frac{\hbar\omega_{\mathbf{q}}}{(\varepsilon_{\mathbf{k}}^0 - \varepsilon_{\mathbf{k-q}}^0)^2 - (\hbar\omega_{\mathbf{q}})^2} N[a_{\mathbf{k},\sigma}^+ a_{\mathbf{k},\sigma}] \tag{7.5}$$

The electron energies $\varepsilon_\mathbf{k}^0$ with the corrections (7.5) represent the well-known quasiparticles – polarons that were originally derived on the basis of Lee-Low-Pines transformation [35]. Now it is clear how the polarons can be directly derived from the general representation as a special solution of the COM problem where the trigger inducing the structural instability is bypassed. Whereas the first part of (7.5) concerns only individual polarons, the general representation yields also the second part of the corrections (7.5), which must be added to the polaron energies. Put differently, every polaron "feels" an effective field of other polarons, ergo, dressed polarons are created.

## 8. Fröhlich Hamiltonian and the BCS theory

We will discuss now the two-particle term (5.9) for conductors. Again, the electron-hyperphonon problem reduces to the electron-phonon one, so after substitution for $c$ and $\tilde{c}$ in (5.11-12) we get:

$$\Delta H_F'' = \sum_{\substack{PQRSr \\ P \neq R, Q \neq S}} u_{PR}^r u_{SQ}^{r*} \frac{\hbar \omega_r [(\varepsilon_P^0 - \varepsilon_R^0)(\varepsilon_S^0 - \varepsilon_Q^0) - (\hbar \omega_r)^2]}{[(\varepsilon_P^0 - \varepsilon_R^0)^2 - (\hbar \omega_r)^2][(\varepsilon_S^0 - \varepsilon_Q^0)^2 - (\hbar \omega_r)^2]} N[a_P^+ a_Q^+ a_S a_R] \qquad (8.1)$$

In solid state notation this term reads ($r \to \mathbf{q}$; $P \to \mathbf{k+q}, \sigma$; $Q \to \mathbf{k'}, \sigma'$; $R \to \mathbf{k}, \sigma$; $S \to \mathbf{k'+q}, \sigma'$):

$$\Delta H_F'' = \sum_{\substack{\mathbf{k}, \mathbf{k'}, \mathbf{q}, \sigma, \sigma' \\ \mathbf{q} \neq 0}} |u^\mathbf{q}|^2 \frac{\hbar \omega_\mathbf{q} [(\varepsilon_{\mathbf{k+q}}^0 - \varepsilon_\mathbf{k}^0)(\varepsilon_{\mathbf{k'}}^0 - \varepsilon_{\mathbf{k'}}^0) - (\hbar \omega_\mathbf{q})^2]}{[(\varepsilon_{\mathbf{k+q}}^0 - \varepsilon_\mathbf{k}^0)^2 - (\hbar \omega_\mathbf{q})^2][(\varepsilon_{\mathbf{k'}}^0 - \varepsilon_{\mathbf{k'}}^0)^2 - (\hbar \omega_\mathbf{q})^2]} N[a_{\mathbf{k+q},\sigma}^+ a_{\mathbf{k'},\sigma'}^+ a_{\mathbf{k'+q},\sigma'} a_{\mathbf{k},\sigma}] \qquad (8.2)$$

When Fröhlich was unsuccessful with his derivation of the ground state energy correction (7.2), regarding the desired gap measured in superconductors, he declared in the last sentence of his second famous paper [10] that the theoretical treatment of superconductivity effects has to wait for the development of new methods for dealing with two-particle effective interaction, based on his transformation. He published it as a challenge that somehow by means of the true many-body treatment, going beyond the Hatree-Fock approximation, the expected gap could be achieved. He gained the following two-particle expression, known as the Fröhlich Hamiltonian:

$$\Delta H_{F(Fr)}'' = \sum_{\substack{\mathbf{k}, \mathbf{k'}, \mathbf{q}, \sigma, \sigma' \\ \mathbf{q} \neq 0}} |u^\mathbf{q}|^2 \frac{\hbar \omega_\mathbf{q}}{(\varepsilon_{\mathbf{k+q}}^0 - \varepsilon_\mathbf{k}^0)^2 - (\hbar \omega_\mathbf{q})^2} a_{\mathbf{k+q},\sigma}^+ a_{\mathbf{k'},\sigma'}^+ a_{\mathbf{k'+q},\sigma'} a_{\mathbf{k},\sigma} \qquad (8.3)$$

Comparing the equations (8.2) and (8.3), we can see that they are different in two details. Our derivation contains the normal product of the creation and annihilation operators; therefore it is the two-particle correction to the one-particle solution represented by selfconsistent polarons (7.5). Fröhlich Hamiltonian does not contain the normal product; it refers directly to electron corrections. But this detail is not important.

The more interesting fact is the difference in the terms containing the electron and vibrational energies caused by application of various transformations (4.1) and (4.2). The first remarkable consequence of this fact is the symmetrical relation between indices $\mathbf{k}$ and $\mathbf{k'}$ in (8.2) that is not fulfilled in the expression (8.3). Wagner was the first who pointed out this problem in the Fröhlich's expression and therefore proposed the effective two-electron interaction gained on the basis of pure adiabatic transformation with the generator $S_l(Q)$ [36]. Later Lenz and Wegner [37] analysed in details the ambiguity of the form of the Fröhlich Hamiltonian by means of the continuous unitary transformations.

This ambiguity problem is also reflected in the reduced form of both Hamiltonian (8.2) and (8.3), used in the BCS theory [11]. Whereas our form of the reduced Hamiltonian is fully attractive,

$$\Delta H_{red} = -2 \sum_{\mathbf{k},\mathbf{k'};\mathbf{k}\neq\mathbf{k'}} |u^{\mathbf{k'}\text{-}\mathbf{k}}|^2 \frac{\hbar\omega_{\mathbf{k'}\text{-}\mathbf{k}}[(\varepsilon_{\mathbf{k'}}^0 - \varepsilon_{\mathbf{k}}^0)^2 + (\hbar\omega_{\mathbf{k'}\text{-}\mathbf{k}})^2]}{[(\varepsilon_{\mathbf{k'}}^0 - \varepsilon_{\mathbf{k}}^0)^2 - (\hbar\omega_{\mathbf{k'}\text{-}\mathbf{k}})^2]^2} a_{\mathbf{k'}\uparrow}^+ a_{\text{-}\mathbf{k'}\downarrow}^+ a_{\text{-}\mathbf{k}\downarrow} a_{\mathbf{k}\uparrow} \qquad (8.4)$$

Fröhlich's reduced Hamiltonian

$$\Delta H_{red(Fr)} = 2 \sum_{\mathbf{k},\mathbf{k'};\mathbf{k}\neq\mathbf{k'}} |u^{\mathbf{k'}\text{-}\mathbf{k}}|^2 \frac{\hbar\omega_{\mathbf{k'}\text{-}\mathbf{k}}}{(\varepsilon_{\mathbf{k'}}^0 - \varepsilon_{\mathbf{k}}^0)^2 - (\hbar\omega_{\mathbf{k'}\text{-}\mathbf{k}})^2} a_{\mathbf{k'}\uparrow}^+ a_{\text{-}\mathbf{k'}\downarrow}^+ a_{\text{-}\mathbf{k}\downarrow} a_{\mathbf{k}\uparrow} \qquad (8.5)$$

has both attractive and repulsive parts.

Although the problem of the correct derivation of the Fröhlich Hamiltonian was deeply discussed in the past, the much more important problem of the possibility of the creation of an energy gap by means of an effective attractive two-electron interaction was never re-examined, in spite of the fact that Fröhlich who first derived this effective two-electron Hamiltonian finally never accepted the two-particle Cooper-pair based theory and claimed that the superconductivity has to be of one-particle origin.

We have studied the influence of two-particle interaction on the removing the degeneracy in continuous spectrum [21] and our results are alarming: This degeneracy can never be removed by a two-particle mechanism. The two-particle mechanism can only decrease the total energy but does not open any gap. It represents only the correlation energy. The detailed analysis was performed in our previous paper [21].

The most important argument against the explanation of superconductivity on the basis of the effective two-electron Hamiltonian follows from the article [22] where Biskupič with his numerical test on $H_2$, HD and $D_2$ molecules confirmed that the Fröhlich based transformations contribute only with cca 20 % of the total value of the adiabatic correlation energy. The error due to the neglecting of the COM problem is 400 % whereas the error of the Hatree-Fock approach is only cca 7 %. This fact implies our most important objection: The role of the COM problem in non-adiabatic cases, as e.g. superconductivity, is much more emergent than the two-particle treatment beyond the Hartree-Fock approach. The true many-body has to be primarily build on the electron-hyperphonon mechanism, and this consequently completely disqualifies the Fröhlich Hamiltonian and all the theories build on it, including the BCS one. They cannot lead to any gap since they describe only the correlation energy of conductors.

## 9. State of superconductivity

Whereas the extended Born-Handy formula (5.14) has the unique solution for small systems (molecules), for the large systems (solids) its solution is ambiguous. We have shown that the solution via bypassing the trigger leads to conductors. Now we will deal with another solution with an active trigger causing the change of the system geometry and removing the electron degeneracy.

Let us consider the conductor with the half-filled conducting band. Rotonic and translonic coupling first splits the initial lattice into two sublattices, so that the new arising system indicates only the half symmetry in respect to the initial one. This implies the splitting of the initial band into two new bands, overlapping on the unperturbed level. We denote the unperturbed energies of the lower valence band as $\varepsilon_{v,k}^0$, and those of the higher conducting band as $\varepsilon_{c,k}^0$. In a similar way we get twice as many hyperphonon branches – innerband with accoustical branches, and interband containing only the optical branches, but moreover rotons and translons. We denote the frequencies of the former set as $\omega_{a,q}$ and the frequencies of the latter set as $\omega_{o,q}$. Finally the vibronic coupling via the optical phonon modes stabilizes the whole system in this new configuration. After the rewriting of the equation (5.14) in solid state notation ($r \rightarrow o, q$; $I \rightarrow v, k, \sigma$; $A \rightarrow c, k', \sigma'$; $\varepsilon_I^0 \rightarrow \varepsilon_{v,k}^0$; $\varepsilon_A^0 \rightarrow \varepsilon_{c,k'}^0$; $u_{AI}^r \rightarrow u_{k'k}^q = u^{k'-k} = u^q$) we get

$$\Delta E_0 = 2 \sum_{k,k'} |u^{k'-k}|^2 \frac{\hbar \omega_{o,k'-k}}{(\varepsilon_{c,k'}^0 - \varepsilon_{v,k}^0)^2 - (\hbar \omega_{o,k'-k})^2}$$

$$+ 4 \sum_{k, r \in R} |u^r|^2 \frac{\rho_r}{(\varepsilon_{c,k}^0 - \varepsilon_{v,k}^0)^2} + 4 \sum_{k, r \in T} |u^r|^2 \frac{\tau_r}{(\varepsilon_{c,k}^0 - \varepsilon_{v,k}^0)^2} \qquad (9.1)$$

The equation (9.1) totally differs from the (7.2) for conductors which was derived by Fröhlich. His equation could never describe superconductors since it supposes only the B-O level of structure typical of conductors. On the other hand, the equation (9.1) fully respects the J-T splitting of bands. All unperturbed energies $\varepsilon_{v,k}^0$ and $\varepsilon_{c,k}^0$ with the same quasimomentum $k$ have to differ in some small nonzero values. Instead of Cooper pairing of two electrons with opposite quasimomenta and spins, as it is stated in the BCS theory, we obtain the pairing between occupied valence and unoccupied conducting band electronic states with the same quasimomenta and spins, i.e. the coherent process over the whole crystal. This leads to the configuration with the single-valued occupancy of states: they are either occupied and belong to the valence band, or are unoccupied and belong to the conducting band. It seams that it is a similar solution, which is typical of insulators or semiconductors. On the contrary, the Fröhlich equation (7.2) leads to the partial occupancy of states and is optimized with respect to the occupation factors, what is typical of conductors, whereas in the equation (9.1) the only optimized parameter is the J-T displacement of the former sublattice with respect to the latter one.

Now we have to answer the question if the optimization process of the ground state energy (9.1) is able to open an energy gap. The diagonal form of the J-T one-particle excitation expression (6.1) is fully justified in solid state physics where the translational symmetry is supposed. Since we have two bands, in solid state notation the one-particle Hamiltonian (6.1) reads

$$\Delta H_F' = \sum_{\mathbf{k},\sigma} \left( \Delta \varepsilon_{v,\mathbf{k}} + \Delta \varepsilon_{c,\mathbf{k}} \right) N[a_{\mathbf{k},\sigma}^+ a_{\mathbf{k},\sigma}] \tag{9.2}$$

so that we have two sets of one-particle corrections, one set for valence band electronic corrections and the latter set for conducting ones.

$$\Delta \varepsilon_{v,\mathbf{k}} = \sum_{\mathbf{q}\neq 0} |u^{\mathbf{q}}|^2 \left( \frac{\hbar\omega_{o,\mathbf{q}}}{(\varepsilon_{v,\mathbf{k}}^0 - \varepsilon_{c,\mathbf{k-q}}^0)^2 - (\hbar\omega_{o,\mathbf{q}})^2} - \frac{\hbar\omega_{a,\mathbf{q}}}{(\varepsilon_{v,\mathbf{k}}^0 - \varepsilon_{v,\mathbf{k-q}}^0)^2 - (\hbar\omega_{a,\mathbf{q}})^2} \right)$$
$$+ 2\sum_{r\in R} |u^r|^2 \frac{\rho_r}{(\varepsilon_{v,\mathbf{k}}^0 - \varepsilon_{c,\mathbf{k}}^0)^2} + 2\sum_{r\in T} |u^r|^2 \frac{\tau_r}{(\varepsilon_{v,\mathbf{k}}^0 - \varepsilon_{c,\mathbf{k}}^0)^2} \tag{9.3}$$

$$\Delta \varepsilon_{c,\mathbf{k}} = -\sum_{\mathbf{q}\neq 0} |u^{\mathbf{q}}|^2 \left( \frac{\hbar\omega_{o,\mathbf{q}}}{(\varepsilon_{c,\mathbf{k}}^0 - \varepsilon_{v,\mathbf{k-q}}^0)^2 - (\hbar\omega_{o,\mathbf{q}})^2} - \frac{\hbar\omega_{a,\mathbf{q}}}{(\varepsilon_{c,\mathbf{k}}^0 - \varepsilon_{c,\mathbf{k-q}}^0)^2 - (\hbar\omega_{a,\mathbf{q}})^2} \right)$$
$$- 2\sum_{r\in R} |u^r|^2 \frac{\rho_r}{(\varepsilon_{c,\mathbf{k}}^0 - \varepsilon_{v,\mathbf{k}}^0)^2} - 2\sum_{r\in T} |u^r|^2 \frac{\tau_r}{(\varepsilon_{c,\mathbf{k}}^0 - \varepsilon_{v,\mathbf{k}}^0)^2} \tag{9.4}$$

We can take notice of innerband frequencies $\omega_{a,\mathbf{q}}$ that are not involved in the ground state energy equation but are present in one-particle correction terms. These terms are the same as those in the reduced Fröhlich's Hamiltonian (8.5), i.e. the denominators of them can achieve both positive and negative values. On the other hand the terms with interband optical frequencies $\omega_{o,\mathbf{q}}$ are optimized by means of the equation (9.1), therefore the negative denominators will be prevailing. This will result in negative values of $\Delta \varepsilon_{v,\mathbf{k}}$ and positive values of $\Delta \varepsilon_{c,\mathbf{k}}$. Of course, from the general form of the equations (9.3-4) we cannot uniquely predicate the existence of a gap. Not all conductors become necessary superconductors at absolute zero. It depends on many factors but the most important factor is the bandwidth. It is apparent from (9.3-4) that the narrow bands (high $T_C$ superconductors) result in greater gaps than broad bands (low $T_C$ superconductors).

The most important fact is that the equations (9.3-4) for the superconducting gap and entirely unlike polaron equations (7.5) for conductors are two different solutions of one common equation (5.8), as well as the ground state equations (9.1) for superconductors and (7.2) for conductors are two solutions of one extended Born-Handy formula (5.7). This strongly contradicts the BCS theory, which seems to be "a better ground state" for conductors.

The privilege position of the extended Born-Handy formula can be seen also in the derivation of the main thermodynamical properties of superconductors. We need not know anything specific about superconductors; the pure assumption of the J-T like solution of this formula is sufficient.

Let us start with the temperature dependent form of the equation (6.4) [13,19].

$$\Delta\varepsilon_P(T) = \sum_{A(T)} \Omega_{PA} - \sum_{I(T)} \Omega_{PI} = \sum_Q \Omega_{PQ}\left(1 - 2f_Q(T)\right) \tag{9.5}$$

Fermions in the general representation naturally obey the Fermi-Dirac statistics and therefore the occupation probability for the state $Q$ is given by the well-known expression

$$f_Q(T) = \frac{1}{e^{\frac{\varepsilon_Q(T)-\mu}{kT}} + 1} \tag{9.6}$$

where $\varepsilon_Q$ is the energy of the fermion state $Q$ (i.e. $\varepsilon_Q^0 + \Delta\varepsilon_Q$). Then the equation (9.5) can be rewritten after substitution from the expression (9.6) as:

$$\Delta\varepsilon_P(T) = \sum_Q \Omega_{PQ}\,\text{tgh}\,\frac{\varepsilon_Q(T)-\mu}{2kT} \tag{9.7}$$

In order to get a reasonable analytical result let us adopt a simplified model where for any virtual state we suppose (in solids this corresponds to an ideal narrow band case):

$$\varepsilon_A(T) - \mu = \Delta\varepsilon(T) \tag{9.8}$$

and for any occupied state:

$$\varepsilon_I(T) - \mu = -\Delta\varepsilon(T) \tag{9.9}$$

Then (9.7) has the form

$$\Delta\varepsilon_P(T) = \Delta\varepsilon_P(0)\,\text{tgh}\,\frac{\Delta\varepsilon(T)}{2kT} \tag{9.10}$$

Further we omit the index $P$ according to the simplifying conditions (9.8-9) and will search for the critical temperature $T_c$ at which the energy gap vanishes. Because the energy gap $\Delta_0$ at the zero temperature is given as:

$$\Delta_0 = 2\Delta\varepsilon(0) \tag{9.11}$$

we finally get the ratio between the energy gap and the critical temperature

$$\frac{\Delta_0}{kT_c} = 4 \tag{9.12}$$

For comparison, in the BCS theory this ratio is 3,52. In relative values both the BCS and our dependence of the energy gap on the temperature are exactly the same (i.e. the dependences of $\Delta/\Delta_0$ on $T/T_c$). The study of other physical properties, such as specific heat, is published in our previous paper [19]. Let us note that the equation (9.12) was derived without any specific requirements for the detailed mechanism of superconductivity in comparison with the BCS theory. It reflects the thermodynamical properties of non-adiabatic systems in a more general form, solely as a consequence of the solution of the extended Born-Handy formula.

As it was mentioned above, the equation (9.1) leads to the ground state, which is distinctive of insulators and semiconductors. How superconductors differ from them? There is one important difference: classical insulators are based on the structure defined by means of the B-O approximation, i.e. the structure with only one real ground state corresponding to the uniquely defined geometry for the minimum total energy of the system. On the other hand, the equation (9.1) is based on the J-T splitting of the original lattice of the conducting state into two sublattices. This splitting is never single-valued but there always exist several (or an infinite number) of equivalent directions of distortion. Therefore we can define superconductors as multigroundstate insulators with several equivalent ground states that correspond to different nuclear positions - Jahn-Teller equivalent configurations.

## 10. Effect of superconductivity

We shall distinguish two fundamental attributes of superconductivity - the state of superconductivity and the effect of superconductivity - that lead to two complementary descriptions of superconductors. On one side the state of superconductivity is characterized by the state of a conducting material, which, after the Jahn-Teller condensation, becomes an insulator with several equivalent ground states. The state of superconductivity determines all statical properties of superconductors: energy gap, its temperature dependence, specific heat, density of states near the Fermi surface etc. On the other side the effect of superconductivity determines all dynamical properties of superconductors: supercurrent, Meissner effect, quantization of magnetic flux, etc. We shall devote in this section just to the problem of effect of superconductivity.

The fact that the superconductor cannot be defined unambiguously on the microscopical level, i.e. that it is characterized by the occurrence of several equivalent groundstates, implies the possibility of spontaneous transition from one ground state into another one. This process, known as the dynamic J-T effect, represents a new degree of freedom of the whole system, which is orthogonal to other degrees of freedom and is also independent on them. It means that this new degree of freedom is quite nondissipative. The transition process has a cooperative long range order property, i.e. the sublattices cannot be deformed (otherwise the conception of two bands would be disturbed) and can only move one with respect to the other. Because the transition from one state into another is conditioned by the overcoming of the potential barrier between two neighbouring ground states we shall speak about the tunnelling process. In this respect we can find a quantum chemical analogy - molecules with two ground states (right torque and left torque). There is also a spontaneous tunnelling transition from one configuration to the other one.

The effect of superconductivity is therefore caused by nuclear microflows through equivalent ground states. There is a question if this nuclear motion and the lattice symmetry lowering can be detectable. Because all the equivalent ground states are symmetrically localized around the symmetrical central point (i.e. the point corresponding to the ground state of material above $T_c$) there are the same probabilities of the occurrence of the system in each of these states. The resulting effect is therefore symmetrical. The experimentally measured nuclear formfactors indicate the rotational ellipsoids originating from the vibrational degrees of freedom. There is a possibility that this new nondissipative „rotational" degree of freedom is hidden in the above mentioned rotational ellipsoids. According to our theory the rotational ellipsoids would be enhanced at the phase transition below $T_c$. And indeed, the recent investigation of structure and superconducting properties of $Nb_3Sn$ ($T_c$=18,5 K) by X-ray diffraction [38] fully confirms the theory presented here. On the studied low-$T_c$ compound $Nb_3Sn$, where the Jahn-Teller effect at the transition from the normal to superconducting state has not been assumed before, a discontinuous increase of the isotopic Debye temperature factors of niobium and tin has been observed in the temperature dependence at cooling near to $T_c$. Maybe the finer experiments show in future some changes in formfactor values of further low- and high-$T_c$ superconductors near the critical temperature.

We have mentioned the state of superconductivity formed by means of the pairing between occupied valence and unoccupied conducting band electronic states with the same quasimomenta and spins, as a consequence of the electron-translon and electron-roton interaction. This is the first pairing process relating to superconductivity. Then we have mentioned the effect of superconductivity caused by the dynamic J-T effect, which on the crystal level with translational symmetry induces the temporary pairing of the neighbouring nuclei. This is the second pairing process. Now the question arises, what is the origin of the superconducting flow of electrons.

It is clear that the dynamic J-T effect affects not only nuclear positions but also the electron distributions. Due to the translational symmetry in crystals this dynamic J-T effect has two levels: on the former level the tunnelling process occurs between the equivalent ground states, causing the movement of two sublattices, and on the latter level the tunnelling process arises between the electron distributions. Therefore we shall speak about the double-level dynamic J-T effect. Whereas the tunnelling of nuclei is limited within the meaning of „there and back", the tunnelling of electron distributions has more abilities - „there and back", „only backwards", and „only forwards". Since the electron distribution of superconductors – multi-ground-state insulators - is always of the closed shell form, the minimum tunnelling electron distribution consists of two electrons with the same quasimomenta and the opposite spins. And this is the third pairing process, which explains the supercurrent with the minimum charge $2e$.

Since the double-level dynamic J-T effect was never investigated before, there is no experience how to treat it exactly. We only know that both levels of this double-level effect induce two new nondissipative degrees of freedom: the former degree for the tunnelling of nuclei (two sublattices) and the latter one for the tunnelling of two-electron pairs. From the preliminary considerations we can only estimate the maximum supercurrent velocity of each electron (i.e. the velocity in the „only forwards" mode). If we denote the frequency of the nuclei relating to the former new degree of freedom as $\omega_N$ and the original full symmetry (i.e. before the J-T splitting) lattice constant as $\mathbf{a}$, the maximum velocity $\mathbf{v}_{max}$ will be defined as

$$\mathbf{v}_{max} = \frac{\omega_N}{2\pi}\mathbf{a} \qquad (10.1)$$

Let us note that only both electrons from each electron pair have to tunnel with the same velocity but the velocities of various pairs are not correlated.

The existence of the latter new degree of freedom has the most important consequence in a fact that the quasimomenta of the tunnelling electron pairs belong to some orthogonal space relating to that one, in which the quasimomenta of electrons in the valence and conducting bands of superconductors – multi-ground-state insulators are defined. Therefore the quasimomenta of the tunnelling pairs cannot be expressed via the $\mathbf{k}$-space any more, but we have to introduce the orthogonal $\mathbf{l}$-space. Each electron from the pair defined via the "statical" quasimomentum $\mathbf{k}$ and spin $\pm\sigma$ moves then as a de Broglie wave with the quasimomentum $\mathbf{l}_\mathbf{k}$. The values of $\mathbf{l}_\mathbf{k}$ are only limited by the maximal value

$$\mathbf{l}_\mathbf{k} \in \langle -\mathbf{l}_{max}, \mathbf{l}_{max} \rangle \qquad (10.2)$$

where the maximal value of $\mathbf{l}_{max}$ can be expressed by means of the equation (10.1):

$$\mathbf{l}_{max} = \frac{\mathbf{p}_{max}}{\hbar} = \frac{m}{\hbar}\mathbf{v}_{max} = \frac{m\omega_N}{h}\mathbf{a} \qquad (10.3)$$

Thus, we have shown that the simple Cooper pair based mechanism cannot explain the origin of superconductivity and that three different pairing mechanisms are necessary to its full understanding. The first and initiating pairing mechanism is related only to electronic states and not to real particles. This type of pairing is responsible for the state of superconductors alias multi-ground-state insulators. Since the excitation mechanism is one-particle, the whole theory describing the state of superconductivity has to be indispensably one-particle. On the other hand, the double-level dynamic J-T effect induces two new nondissipative degrees of freedom accompanying with the pairing of real particles during the tunnelling process – temporary pairing of neighbouring nuclei and the pairing of two electrons with the same quasimomenta and opposite spins. This is the final effect o superconductivity described on two-particle basis.

The above mentioned conclusions influence the concept of correspondence between macrostates and microstates. It is commonly believed that any macrostate of superconductor with a certain value of supercurrent corresponds to one appropriate microstate described by a certain value of charge carrier quasimomentum. According to our theory the macrostate with zero supercurrent corresponds to several microstates, i.e. microscopical configurations representing equivalent ground states, and any other macrostate with nonzero supercurrent corresponds to a certain transition process between these microscopical configurations.

Further we mention the conception of two phases: superconducting and conducting. This conception originates from the phenomenological idea of parallel coexistence of two phase components - superconducting (x) and conducting (1-x). It is motivated by the classical thermodynamics where in a similar way e.g. the coexistence of liquid and gaseous phases of the same matter is described. This macroscopical phenomenological conception was later incorporated in microscopical theories. So, in compliance with the BCS theory, the Cooper-paired electrons representing the superconducting phase coexist with free non-paired electrons representing the conducting phase in a parallel way. On the contrary to this our theory considers these two phases to be not parallel but orthogonal in the ontological sense. What does this important difference mean?

In the two-particle theories based on the Cooper pair idea two different entities are identified: the entity responsible for the condensation and excitation mechanism leading to the gap formation and the entity responsible for the transfer of supercurrent. Cooper pairs are the Bose condensation, which decay into free conducting electrons through the excitation mechanism, and simultaneously they are carriers of superconducting current.

In our theory we sharply distinguish these two entities. The former one corresponds to the one-electron J-T excitations. The condensation process represents the creation of the multi-ground-state insulator with fully occupied valence band and empty conducting band. The excitation mechanism is one-particle in principle. Conducting phase of superconductor in this sense resembles the conductance of thermally excited insulator (semiconductor). The condensation and excitation mechanism is a subject of investigation of the state of superconductivity.

The latter entity corresponds to the tunnelling of two-electron distributions (in the delocalised terminology) or double occupied binding orbitals (in the localised terminology), which are the carriers of supercurrent. By this process one set of paired nuclei decays and another one arises. The tunnelling process is two-particle in principle, is connected with two new nondissipative degrees of freedom, one for sublattices and one for paired electrons, and is orthogonal with respect to the electron-hyperphonon interaction mechanism, which is responsible for the one-particle gap formation. The carriers of supercurrent are subject of investigation of the effect of superconductivity.

## 11. Conclusion

The main goal of this work was the implementation of the COM problem into the many-body treatment. The many years experience with the inconvenience of the direct COM separation on the molecular level and its consequent replacement with the Born-Handy ansatz as a full equivalent was taken into account. It was shown that the many-body treatment based on the electron-vibrational Hamiltonian is fundamentally inconsistent with the Born-Handy ansatz so that such a treatment can never respect the COM problem.

The only way-out insists in the requirement, to take into account the whole electron-vibration-rotation-translational Hamiltonian. It means, that the total Hamiltonian in the crude representation, expressed in the second quantization formalism, has explicitly contain not only the vibrational energy quanta, but also the rotational and translational ones, which originate from the kinetic secular matrix. We shall call these new quasiparticles - rotational and translational quanta - as rotons and translons, in a full analogy with phonons in solid state physics. This is a background of the true many-body treatment in quantum chemistry and solid state physics, which we shall call COM many-body theory.

The quasiparticle transformations, binding individual representations of the total Hamiltonian, are then the generalization of the original Fröhlich transformations in such a way that they contain besides the electron-vibrational (vibronic or electron-phonon) interaction moreover the electron-rotational (rotonic or electron-roton) and the electron-translational (translonic or electron-translon) interactions. In order to achieve the unique covariant description of all equations with respect to individual degrees of freedom, we introduce the concept of hypervibrations (hyperphonons), i.e. vibrations + rotations + translations together, and the consequent concept of electron-hypervibrational (hypervibronic or electron-hyperphonon) interaction. We have proved that due to the COM problem only the hypervibrations (hyperphonons) have true physical meaning in molecules and crystals; nevertheless, the use of pure vibrations (phonons) is justified only in the adiabatic systems, i.e. the case when electron energies are much greater the vibrational ones. This fact calls for the total revision of our contemporary knowledge of all non-adiabatic systems.

The most important equation, derived in this work, is the extended Born-Handy formula, valid in the adiabatic limit as well as in the case of break down of B-O approximation. Since due to many-body formulation the extended Born-Handy formula can be expressed in the CPHF compatible form, the extended CPHF equations, describing the non-adiabatic systems, immediately follow from the presented theory. We shall call them COM CPHF equations. Whereas in the adiabatic limit the extended Born-Handy formula represents only small corrections to the system total energy, in non-adiabatic systems it plays three important roles: 1) removes the electron degeneracies, 2) is responsible for the symmetry breaking, and 3) forms the molecular and crystallic structure.

The first role - removing of electron degeneracies - is fulfilled via the vibronic coupling. The second role - symmetry breaking - is caused by the rotonic and translonic coupling. Finally the third role - forming of structure - is a result of optimalization where all three types of coupling participate. Only in the adiabatic limit the forming of molecular and crystallic structure reduces to the standard one, defined by B-O approximation. Moreover, at finite temperatures the extended Born-Handy formula plays yet another role: it defines all thermodynamic properties of the non-adiabatic systems, as was demonstrated on the derivation of the critical temperature of superconductors.

Since the J-T effect was always studied without the inclusion of the COM problem, only vibronic coupling was taken into account, and therefore the symmetry breaking and forming of structure were misunderstood. The trigger causing the system instability has the origin in rotonic and translonic coupling. It is necessary to reformulate the J-T theorem in a new way. One possible formulation is proposed here: "Molecular and crystallic entities in a geometry of electronically degenerate ground state are unstable at this geometry except the case when all matrix elements of electron-rotational and electron-translational interaction equal zero."

As it was mentioned in the introduction, the modern attempts to explain HTSC by means of J-T approach operate with the term "strong vibronic coupling" in order to advocate the presence of the J-T effect, and on the other hand they tolerate the BCS theory for LTSC where only the "week vibronic coupling" occurs, so that the J-T effect may be ignored. But after inclusion of the COM problem we come to conclusion that the question of strong or week vibronic coupling is absolutely irrelevant for the applicability of the J-T effect. Only the symmetry breaking, stimulated by the rotonic and translonic coupling, and acting as a trigger, plays the decisive role. Either the trigger is bypassed, and then the crystal remains in conducting state; or it is switched on and the J-T effect is active, and this leads to superconducting state.

It is the fundamental problem of solid state physics, that due to the false many-body treatment the true nature of crystallic structure beyond the B-O approximation was never revealed. This is the reason why the BCS theory, in spite of the fact that Fröhlich was critical to it and disregarded it as an ontological nonsense, survives more than a half of century till now. The BCS theory is based on the naive belief that the structure of superconductors is the same as the structure of conductors, i.e. that it is defined through the B-O approximation. As we have shown in our previous work [21], there is no mechanism, which could split the degenerate electronic spectrum of conductors and open an energy gap on the adiabatic level.

Fröhlich applied the unitary transformation on the Hamiltonian describing conductors, but his attempt to remove degeneracy failed. Then he proposed the "true" many body treatment. Bardeen with Cooper and Schrieffer continued to fulfil the Fröhlich's idea, and with the full multiconfiguration method used on the Fröhlich's transformed Hamiltonian, they tried to remove the degeneracy. After the two-year intensive work they had no positive solution. At the last moment Bardeen accepted the trial function proposed by Schrieffer, inconsistent with the particle conservation law, and leading to the concept of two-particle Cooper pair based theory, known as BCS. Nevertheless, the solid state Hamiltonian does not contain the information of superconductivity, so that Fröhlich as well as Bardeen calculated only the correlation energy of conductors unless being aware of this important fact.

When the quantum field many body techniques are not transferable into the quantum chemistry dealing with small systems (molecules), they are not fully transferable into the solid state physics dealing with great systems (crystals), too. As it was explained in details in this paper, only the COM many body is the true many body, and is namely inevitable in non-adiabatic cases. For the non-adiabatic crystals the state of conductivity and superconductivity are two possible solutions of the extended Born-Handy formula. This is a quite different view from that using only the classical many body (without COM). The non-adiabatic crystals lead always to system splitting in two subsystems. In the case of conductors the first subsystem is the "adiabatic core" consisting of nuclei and all valence bands, and the second subsystem is the "fluid" of quasi-free conducting electrons. The explanation of conductors on the basis of COM true many body is not so simple as in the case of the classical many body. Whereas for the classical many body the conducting state is a real ground state (and superconductors are something like a "better ground state" after the Bose condensation of Cooper pairs), the COM many body presents conducting state as an abnormal solution of the extended Born-Handy formula, as some excited state, representing the crystal analogy of ionized molecular systems where the ionized electrons are not the part of the molecular "ion core" as well. In other words, the conductors "fling away" one set of electrons so that the rest - the "core", consisting of nuclei and valence bands, is not degenerate any more and is therefore adiabatic with the well-defined structure on the basis of the B-O approximation. The "off-cast" electrons are not the integral part of the adiabatic "core" anymore, but belong still to the system and interact with the "core" via the standard electron-phonon interaction.

In spite of the fact that the effective Hamiltonian of the type adiabatic core + electronic fluid + electron-phonon interaction between them describes conductors correctly, the origin of conductivity was misunderstood, and consequently, the superconductivity was misunderstood as well. The state of superconductivity is not a "better ground state" as the BCS theory explains it, but the real ground state, and moreover, the multi ground state due to the J-T effect. This is the only difference between superconductors and insulators: whereas the latter are simple ground state, the former are multi ground state. Electron-phonon mechanism can never describe superconductivity; we need the complete COM many body, i.e. the electron-hyperphonon mechanism. The rotonic a translonic coupling splits the system into two subsystems - two sublattices, causes the symmetry lowering, defines new non-adiabatic structure, and creates the pairs from all occupied valence and unoccupied conducting band electronic states with the same quasimomenta and spins, with the cooperative behaviour over the whole crystal. This is only a pairing of states and not of real particles; therefore the theory of superconductivity is one-particle in principle, as Fröhlich demanded. Finally the vibronic coupling via the optical phonon modes stabilizes the whole system in this new configuration and opens an energy gap. No set of electrons is "flung away" as in the case of conductors, all electrons are located due the symmetry lowering in the fully occupied valence bands, exactly as in the case of insulators.

We shall distinguish two fundamental attributes of superconductivity - the state and the effect of superconductivity - that lead to two complementary descriptions of superconductors. On one side the state of superconductivity is characterized by the state of a conducting material, which, after the Jahn-Teller condensation, becomes an insulator with several equivalent ground states. The state of superconductivity determines all statical properties of superconductors: energy gap, its temperature dependence, specific heat, density of states near the Fermi surface etc. On the other side the effect of superconductivity determines all dynamical properties of superconductors: supercurrent, Meissner effect, quantization of magnetic flux, etc. Whereas the state of superconductivity is of one-particle nature, the effect is two-particle in principle. The multi ground state character of superconductors leads to the double-level dynamic J-T effect, which induces two new nondissipative degrees of freedom accompanying with the pairing of real particles during the tunnelling process - temporary pairing of neighbouring nuclei and the pairing of two electrons with the same quasimomenta and opposite spins. The new degree of freedom for electrons implies the new quasimomentum space orthogonal to that one where the electrons are described in the state of superconductor - multi ground state insulator. This is the final effect of superconductivity described on two-particle basis, which explains the supercurrent with the minimum charge 2e.

## 12. Appendix A

We present here some useful relations for the expansions of equations (3.7-10) only up to the second order of the Taylor expansion since we will not take the anharmonic terms into account in this paper. Maybe the following relations are trivial but it is worth to mention them due to the cross-platform notation.

$$h_{PQ}^0 = h_{QP}^{0*} \qquad v_{PQRS}^0 = v_{QPSR}^0 = v_{SRQP}^{0\ *} = v_{RSPQ}^{0\ *} \qquad (A1)$$

$$E_{NN}^r = E_{NN}^{\hat{r}\ *} \qquad E_{NN}^{rs} = E_{NN}^{sr} = E_{NN}^{\bar{r}\bar{s}\ *} = E_{NN}^{\hat{s}\hat{r}\ *} \qquad (A2)$$

$$u_{PQ}^r = u_{QP}^{\hat{r}\ *} \qquad u_{PQ}^{rs} = u_{PQ}^{sr} = u_{QP}^{\bar{r}\bar{s}\ *} = u_{QP}^{\hat{s}\hat{r}\ *} \qquad (A3)$$

Let us suppose that for every spinorbital $X$ there exists a unique spinorbital $\hat{X}$ so that the following symmetrical relations hold:

$$h_{PQ}^0 = h_{\hat{Q}\hat{P}}^0 \qquad v_{PQRS}^0 = v_{\hat{R}\hat{S}\hat{P}\hat{Q}}^0 \qquad u_{PQ}^r = u_{\hat{Q}\hat{P}}^r \qquad (A4)$$

In quantum chemistry where the real wave functions are used and the identity $X = \hat{X}$ is supposed, the equations (A4) hold trivially. In solid state theory where the assignment $X \to \mathbf{k}, \sigma$ and $\hat{X} \to -\mathbf{k}, \pm\sigma$ ($\mathbf{k}$ is the electron quasimomentum and $\sigma$ is the spin) is done, these equations lead to the well-known symmetrical relations.

In a similar way, we can expand up to the second order the unitary conditions (4.5-6) of the adiabatic transformation

$$c_{PQ}^0 = \delta_{PQ} \qquad c_{PQ}^r + c_{QP}^{\check{r}*} = 0 \qquad c_{PQ}^{rs} + c_{QP}^{\check{r}\check{s}*} = -\sum_R \left( c_{PR}^r c_{QR}^{\check{s}*} + c_{PR}^s c_{QR}^{\check{r}*} \right) \tag{A5}$$

$$d_{rPQ}^0 = c_{PQ}^{\check{r}} \qquad d_{rPQ}^s = c_{PQ}^{\check{r}s} + \sum_R c_{RP}^{\check{s}*} c_{RQ}^{\check{r}} \tag{A6}$$

and the unitary conditions (4.9-10) for the nonadiabatic transformation

$$\tilde{c}_{PQ}^0 = \delta_{PQ} \qquad \tilde{c}_{PQ}^r - \tilde{c}_{QP}^{\check{r}*} = 0 \qquad \tilde{c}_{PQ}^{rs} + \tilde{c}_{QP}^{\check{r}\check{s}*} = \sum_R \left( \tilde{c}_{PR}^r \tilde{c}_{QR}^{\check{s}*} + \tilde{c}_{PR}^s \tilde{c}_{QR}^{\check{r}*} \right) \tag{A7}$$

$$\tilde{d}_{rPQ}^0 = -\tilde{c}_{PQ}^{\check{r}} \qquad \tilde{d}_{rPQ}^s = -\tilde{c}_{PQ}^{\check{r}s} + \sum_R \tilde{c}_{RP}^{\check{s}*} \tilde{c}_{RQ}^{\check{r}} \tag{A8}$$

## 13. Appendix B

The terms of the electron-hypervibrational Hamiltonian up to the second order of Taylor expansion in the general representation are presented here in details. The notation $H_X^{n(k,l)}$ is used where $X$ denotes the terms originating from transformation of the part $A$ or $B$ of the crude Hamiltonian, $n$ represents the power of the Taylor expansion, $k$ the power of the coordinate operator $B_r$ and $l$ the power of the momentum operator $\tilde{B}_r$.

$$H_A^0 = E_{NN}^0 + \sum_{PQ} h_{PQ}^0 a_P^+ a_Q + \tfrac{1}{2} \sum_{PQRS} v_{PQRS}^0 a_P^+ a_Q^+ a_S a_R \tag{B1}$$

$$H_A^{1(1,0)} = \sum_r E_{NN}^r B_r + \sum_{PQr} \left[ u_{PQ}^r + \sum_r \left( h_{PR}^0 c_{RQ}^r + h_{RQ}^0 c_{RP}^{\check{r}*} \right) \right] B_r a_P^+ a_Q$$
$$+ \sum_{PQRST r} \left( v_{PQTS}^0 c_{TR}^r + v_{TQRS}^0 c_{TP}^{\check{r}*} \right) B_r a_P^+ a_Q^+ a_S a_R \tag{B2}$$

$$H_A^{1(0,1)} = \sum_{PQRr} \left( h_{PR}^0 \tilde{c}_{RQ}^r - h_{RQ}^0 \tilde{c}_{RP}^{\check{r}*} \right) \tilde{B}_r a_P^+ a_Q + \sum_{PQRST r} \left( v_{PQTS}^0 \tilde{c}_{TR}^r - v_{TQRS}^0 \tilde{c}_{TP}^{\check{r}*} \right) \tilde{B}_r a_P^+ a_Q^+ a_S a_R \tag{B3}$$

$$H_A^{2(2,0)} = -\tfrac{1}{2} \sum_{rs} V_N^{rs} B_r B_s + \sum_{PQrs} \left[ \tfrac{1}{2} u_{PQ}^{rs} + \sum_R \left( \tfrac{1}{2} h_{PR}^0 c_{RQ}^{rs} + \tfrac{1}{2} h_{RQ}^0 c_{RP}^{\check{r}\check{s}*} + u_{PR}^r c_{RQ}^s + u_{RQ}^r c_{RP}^{\check{s}*} \right) \right.$$
$$\left. + \sum_{RS} h_{RS}^0 c_{RP}^{\check{r}*} c_{SQ}^s \right] B_r B_s a_P^+ a_Q + \tfrac{1}{2} \sum_{PQRST rs} \left\{ v_{PQTS}^0 c_{TR}^{rs} + v_{TQRS}^0 c_{TP}^{\check{r}\check{s}*} + \sum_U \left[ v_{PQTU}^0 c_{TR}^r c_{US}^s \right. \right.$$
$$\left. \left. + v_{TURS}^0 c_{TP}^{\check{r}*} c_{UQ}^{\check{s}*} + 2 \left( v_{TQUS}^0 - v_{TQSU}^0 \right) c_{TP}^{\check{r}*} c_{UR}^s \right] \right\} B_r B_s a_P^+ a_Q^+ a_S a_R \tag{B4}$$

$$H_A^{2(0,2)} = -\tfrac{1}{2} \sum_{rs} W_N^{rs} \tilde{B}_r \tilde{B}_s + \sum_{PQRrs} \left( \tfrac{1}{2} h_{PR}^0 \tilde{c}_{RQ}^{rs} + \tfrac{1}{2} h_{RQ}^0 \tilde{c}_{RP}^{\check{r}\check{s}*} - \sum_S h_{RS}^0 \tilde{c}_{RP}^{\check{r}*} \tilde{c}_{SQ}^s \right) \tilde{B}_r \tilde{B}_s a_P^+ a_Q$$
$$+ \tfrac{1}{2} \sum_{PQRST rs} \left\{ v_{PQTS}^0 \tilde{c}_{TR}^{rs} + v_{TQRS}^0 \tilde{c}_{TP}^{\check{r}\check{s}*} + \sum_U \left[ v_{PQTU}^0 \tilde{c}_{TR}^r \tilde{c}_{US}^s + v_{TURS}^0 \tilde{c}_{TP}^{\check{r}*} \tilde{c}_{UQ}^{\check{s}*} \right. \right.$$
$$\left. \left. + 2 \left( v_{TQSU}^0 - v_{TQUS}^0 \right) \tilde{c}_{TP}^{\check{r}*} \tilde{c}_{UR}^s \right] \right\} \tilde{B}_r \tilde{B}_s a_P^+ a_Q^+ a_S a_R \tag{B5}$$

$$H_A^{2(1,1)} = \frac{1}{2} \sum_{PQRrs} \left\{ u_{PR}^r \tilde{c}_{RQ}^s - u_{RQ}^r \tilde{c}_{RP}^{s*} + \sum_S \left[ \left( h_{PR}^0 c_{RS}^r + h_{RS}^0 c_{RP}^{r*} \right) \tilde{c}_{SQ}^s - \left( h_{SR}^0 c_{RQ}^r + h_{RQ}^0 c_{RS}^{r*} \right) \tilde{c}_{SP}^{s*} \right] \right\}$$

$$. \left( B_r \tilde{B}_s + \tilde{B}_s B_r \right) a_P^+ a_Q + \frac{1}{2} \sum_{PQRSTUrs} \left[ \left( v_{PQTS}^0 c_{TU}^r - v_{PQTU}^0 c_{TS}^r \right) \tilde{c}_{UQ}^s + \left( v_{TURS}^0 c_{TQ}^{r*} - v_{TQRS}^0 c_{TU}^{r*} \right) \tilde{c}_{UP}^{s*} \right.$$

$$\left. + \left( v_{TQUS}^0 - v_{TQSU}^0 \right) \left( c_{TP}^{r*} \tilde{c}_{UR}^s - c_{UR}^r \tilde{c}_{TP}^{s*} \right) \right] \left( B_r \tilde{B}_s + \tilde{B}_s B_r \right) a_P^+ a_Q^+ a_S a_R \quad \text{(B6)}$$

$$H_A^{2(0,0)} = 2 \sum_{PQ} E_{NN}^r \tilde{d}_{rPQ}^0 a_P^+ a_Q + \sum_{PQRr} \left\{ \left[ u_{PR}^r + \sum_S \left( h_{PS}^0 c_{SR}^r + h_{SR}^0 c_{SP}^{r*} \right) \right] \tilde{d}_{rRQ}^0 + \left[ u_{RQ}^r \right. \right.$$

$$\left. \left. + \sum_S \left( h_{RS}^0 c_{SQ}^r + h_{SQ}^0 c_{SR}^{r*} \right) \right] \tilde{d}_{rPR}^0 \right\} a_P^+ a_Q + 2 \sum_{PQRSr} \left[ u_{PR}^r + \sum_T \left( h_{PT}^0 c_{TR}^r + h_{TR}^0 c_{TP}^{r*} \right) \right] \tilde{d}_{rQS}^0 a_P^+ a_Q^+ a_S a_R$$

$$+ \sum_{PQRSTUr} \left[ \left( v_{PQTS}^0 c_{TU}^r - v_{PQTU}^0 c_{TS}^r \right) \tilde{d}_{rUR}^0 + \left( v_{TQRS}^0 c_{TU}^{r*} - v_{TURS}^0 c_{TQ}^{r*} \right) \tilde{d}_{rPU}^0 + \left( v_{TQUS}^0 - v_{TQSU}^0 \right) \right.$$

$$\left. . \left( c_{TP}^{r*} \tilde{d}_{rUR}^0 + c_{UR}^r \tilde{d}_{rPT}^0 \right) \right] a_P^+ a_Q^+ a_S a_R + 2 \sum_{PQRSTUVr} \left( v_{PQVT}^0 c_{VS}^r + v_{VQST}^0 c_{VP}^{r*} \right) \tilde{d}_{rRU}^0 a_P^+ a_Q^+ a_R^+ a_U a_T a_S \quad \text{(B7)}$$

$$H_B^0 = \frac{1}{4} \sum_r \left( \hbar \omega_r B_r^+ B_r + \hbar \tilde{\omega}_r \tilde{B}_r^+ \tilde{B}_r \right) = \frac{1}{4} \sum_r \left( \hbar \omega_r B_r B_r - \hbar \tilde{\omega}_r \tilde{B}_r \tilde{B}_r \right) \quad \text{(B8)}$$

$$H_B^{1(1,0)} = \sum_{PQr} \hbar \omega_r \tilde{d}_{rPQ}^0 B_r a_P^+ a_Q \quad \text{(B9)}$$

$$H_B^{1(0,1)} = -\sum_{PQr} \hbar \tilde{\omega}_r d_{rPQ}^0 \tilde{B}_r a_P^+ a_Q \quad \text{(B10)}$$

$$H_B^{2(2,0)} = 0 \quad \text{(B11)}$$

$$H_B^{2(0,2)} = -\sum_{PQRrs} \hbar \tilde{\omega}_r \left( d_{rPR}^0 \tilde{c}_{RQ}^s - d_{rRQ}^0 \tilde{c}_{RP}^{s*} \right) \tilde{B}_r \tilde{B}_s a_P^+ a_Q \quad \text{(B12)}$$

$$H_B^{2(1,1)} = \frac{1}{2} \sum_{PQrs} \left( \hbar \omega_r \tilde{d}_{rPQ}^s - \hbar \tilde{\omega}_s d_{sPQ}^r \right) \left( B_r \tilde{B}_s + \tilde{B}_s B_r \right) a_P^+ a_Q \quad \text{(B13)}$$

$$H_B^{2(0,0)} = \sum_{PQRr} \left( \hbar \omega_r \tilde{d}_{rPR}^0 \tilde{d}_{rRQ}^0 - \hbar \tilde{\omega}_r d_{rPR}^0 d_{rRQ}^0 \right) a_P^+ a_Q$$

$$+ \sum_{PQRSr} \left( \hbar \omega_r \tilde{d}_{rPR}^0 \tilde{d}_{rQS}^0 - \hbar \tilde{\omega}_r d_{rPR}^0 d_{rQS}^0 \right) a_P^+ a_Q^+ a_S a_R \quad \text{(B14)}$$

# 14. Appendix C

Here is the detailed derivation of the fermion part of the Hamiltonian in the general representation. $\Delta H_F$ from the equation (5.1) can be expressed as a sum of two contributions

$$\Delta H_F = \Delta H_{[F]} + \Delta H_{\langle F \rangle} \quad \text{(C1)}$$

where $\Delta H_{[F]}$ is of a pure fermion origin, i.e. the part that is invariant against boson (hypervibrational) excitations, and $\Delta H_{\langle F \rangle}$ represents the effective part dependent on boson excitations. We take into account only the boson vacuum mean values:

$$\langle 0 | B_r | 0 \rangle = \langle 0 | \tilde{B}_r | 0 \rangle = \langle 0 | B_r \tilde{B}_s + \tilde{B}_s B_r | 0 \rangle = 0 \quad \text{(C2)}$$

$$\langle 0|B_r B_s|0\rangle = -\langle 0|\tilde{B}_r \tilde{B}_s|0\rangle = \delta_{rs} \tag{C3}$$

If we introduce the discrete particle-hole occupation factors

$$h(A) = 0, \ h(I) = 1, \ p(A) = 1, \ p(I) = 0 \tag{C4}$$

and a simplifying notation for the adiabatic derivatives $v^r_{PQRS}$ of the coulomb interaction $v^0_{PQRS}$

$$v^r_{PQRS} = \sum_T \left( v^0_{PQTS} c^r_{TR} + v^0_{PQRT} c^r_{TS} - v^0_{TQRS} c^r_{PT} - v^0_{PTRS} c^r_{QT} \right) \tag{C5}$$

we finally get the individual contributions for one- and two-fermion terms. The surprising fact is the occurrence of three-fermion term, in spite of the fact the crude representation contains only one- and two-electron terms.

$$\Delta H'_{[F]} = \sum_{PQr} \left[ \hbar\tilde{\omega}_r \left( \sum_A c^r_{PA} c^{r*}_{QA} - \sum_I c^r_{PI} c^{r*}_{QI} \right) - \hbar\omega_r \left( \sum_A \tilde{c}^r_{PA} \tilde{c}^{r*}_{QA} - \sum_I \tilde{c}^r_{PI} \tilde{c}^{r*}_{QI} \right) \right] N[a^+_P a_Q]$$

$$-2\sum_{PQr} E^{r*} \tilde{c}^r_{PQ} N[a^+_P a_Q] + \sum_{PQr} \left[ \left( h(P) - p(P) \right)\varepsilon^{r*}_P + \left( h(Q) - p(Q) \right)\varepsilon^{r*}_Q \right] \tilde{c}^r_{PQ} N[a^+_P a_Q]$$

$$- \sum_{PQAIr} \left[ \left( v^r_{PIQA} - v^r_{PIAQ} \right)\tilde{c}^{r*}_{IA} + \left( v^r_{PAQI} - v^r_{PAIQ} \right)\tilde{c}^{r*}_{AI} \right] N[a^+_P a_Q] \tag{C6}$$

$$\Delta H''_{[F]} = \frac{1}{2} \sum_{PQRS} v^0_{PQRS} N[a^+_P a^+_Q a_S a_R] + \sum_{PQRSr} \left( \hbar\tilde{\omega}_r c^r_{PR} c^{r*}_{SQ} - \hbar\omega_r \tilde{c}^r_{PR} \tilde{c}^{r*}_{SQ} \right) N[a^+_P a^+_Q a_S a_R]$$

$$-2\sum_{PQSr} \varepsilon^r_P \tilde{c}^{r*}_{SQ} N[a^+_P a^+_Q a_S a_P] + \sum_{PQRSTr} \left\{ \sum_I \left[ v^0_{PQTS} c^r_{TI} - v^0_{PQTI} c^r_{TS} + \left( v^0_{TQSI} - v^0_{TQIS} \right) c^r_{PT} \right] \tilde{c}^{r*}_{RI} \right.$$

$$+ \sum_I \left[ v^0_{TIRS} c^r_{QT} - v^0_{TQRS} c^r_{IT} + \left( v^0_{IQTS} - v^0_{IQST} \right) c^r_{TR} \right] \tilde{c}^{r*}_{IP}$$

$$- \sum_A \left[ v^0_{PQTS} c^r_{TA} - v^0_{PQTA} c^r_{TS} + \left( v^0_{TQSA} - v^0_{TQAS} \right) c^r_{PT} \right] \tilde{c}^{r*}_{RA}$$

$$\left. - \sum_A \left[ v^0_{TARS} c^r_{QT} - v^0_{TQRS} c^r_{AT} + \left( v^0_{AQTS} - v^0_{AQST} \right) c^r_{TR} \right] \tilde{c}^{r*}_{AP} \right\} N[a^+_P a^+_Q a_S a_R] \tag{C7}$$

$$\Delta H'''_{[F]} = -2 \sum_{PQRSTUVr} \left( v^0_{PQVT} c^r_{VS} - v^0_{VQST} c^r_{PV} \right) \tilde{c}^{r*}_{UR} N[a^+_P a^+_Q a^+_R a_U a_T a_S] \tag{C8}$$

$$H'_{\langle F\rangle} = \frac{1}{2} \sum_{Pr} u^{rr}_{PP} N[a^+_P a_P] + \sum_{PRr} \left[ \left( \varepsilon^0_P - \varepsilon^0_R \right) \left( |c^r_{PR}|^2 + |\tilde{c}^r_{PR}|^2 \right) - 2\hbar\tilde{\omega}_r \, \text{Re}\left( \tilde{c}^r_{PR} c^{r*}_{PR} \right) \right] N[a^+_P a_P]$$

$$+ \sum_{PAIr} \left\{ \sum_R \left[ \left( v^0_{PRPA} - v^0_{PRAP} \right) \left( c^r_{IR} c^{r*}_{IA} + \tilde{c}^r_{IR} \tilde{c}^{r*}_{IA} \right) - \left( v^0_{PRPI} - v^0_{PRIP} \right) \left( c^r_{AR} c^{r*}_{AI} + \tilde{c}^r_{AR} \tilde{c}^{r*}_{AI} \right) \right] \right.$$

$$\left. + \frac{1}{2} \left[ \left( v^0_{PIPA} - v^0_{PIAP} \right) \left( c^{rr}_{AI} - \tilde{c}^{rr}_{AI} \right) - \left( v^0_{PAPI} - v^0_{PAIP} \right) \left( c^{rr}_{IA} - \tilde{c}^{rr}_{IA} \right) \right] \right\} N[a^+_P a_P] \tag{C9}$$

$$H''_{\langle F\rangle} = \frac{1}{2} \sum_{PQRSTr} \left\{ v^0_{PQTS} \left( c^{rr}_{TR} - \tilde{c}^{rr}_{TR} \right) + v^0_{TQRS} \left( c^{rr*}_{TP} - \tilde{c}^{rr*}_{TP} \right) - \sum_U \left[ v^0_{PQTU} \left( c^r_{TR} c^{r*}_{SU} + \tilde{c}^r_{TR} \tilde{c}^{r*}_{SU} \right) \right. \right.$$

$$\left. \left. + v^0_{TURS} \left( c^r_{PT} c^{r*}_{UQ} + \tilde{c}^r_{PT} \tilde{c}^{r*}_{UQ} \right) + 2\left( v^0_{TQSU} - v^0_{TQUS} \right) \left( c^r_{PT} c^{r*}_{RU} + \tilde{c}^r_{PT} \tilde{c}^{r*}_{RU} \right) \right] \right\} N[a^+_P a^+_Q a_S a_R] \tag{C10}$$

## Acknowledgements

The author wishes to express his gratitude to E. Brändas for his valuable advice during compilation of this paper, to O. Šipr for critical reading of the manuscript and useful suggestions and to V. Žárský for constant help and encouragement.